\documentclass[american,prl,twocolumn]{revtex4-1}
\usepackage[T1]{fontenc}
\usepackage[latin9]{inputenc}
\usepackage{booktabs}
\usepackage{amsmath}
\usepackage{graphicx}

\makeatletter

\providecommand{\tabularnewline}{\\}

\makeatother

\usepackage{babel}
\begin{document}

\title{A nested polyhedra model of turbulence}

\author{Ö. D. Gürcan$^{1,2}$}

\affiliation{$^{1}$ CNRS, Laboratoire de Physique des Plasmas, Ecole Polytechnique,
Palaiseau}

\affiliation{$^{2}$ Sorbonne Universités, UPMC Univ Paris 06, Paris}
\begin{abstract}
A discretization of the wave-number space is proposed, using nested
polyhedra, in the form of alternating dodecahedra and icosahedra that
are self-similarly scaled. This particular choice allows the possibility
of forming triangles using only discretized wave-vectors when the
scaling between two consecutive dodecahedra is equal to the \emph{golden
ratio}, and the icosahedron between the two dodecahedra is the dual
of the inner dodecahedron. Alternatively, the same discretization
can be described as a logarithmically spaced (with a scaling equal
to the golden ratio), nested dodecahedron-icosahedron compounds. A
wave-vector which points from the origin to a vertex of such a mesh,
can always find two other discretized wave-vectors that are also on
the vertices of the mesh (which is not true for an arbitrary mesh).
Thus, the nested polyhedra grid can be thought of as a reduction (or
decimation) of the Fourier space using a particular set of self-similar
triads aranged approximately in a spherical form. For each vertex
(i.e. discretized wave-vector) in this space, there are either 9 or
15 pairs of vertices (i.e. wave-vectors) with which the initial vertex
can interact to form a triangle. This allows the reduction of the
convolution integral in the Navier-Stokes equation to a sum over 9
or 15 interaction pairs. Transforming the equation in Fourier space,
to a network of ``interacting'' nodes, that can be constructed as
a numerical model, which evolves each component of the velocity vector
on each node of the network. Such a model gives the usual Kolmogorov
spectrum of $k^{-5/3}$. Since the scaling is logarithmic, and the
number of nodes for each scale is constant, a very large inertial
range (i.e. a very high Reynolds number) can be considered with a
much lower number of degrees of freedom. Incidentally, by assuming
isotropy and a certain relation between the phases, the model can
be used to systematically derive shell models.

\end{abstract}
\maketitle

\section{Introduction}

Turbulence is a complex phenomenon involving chaotic behaviour over
a range of scales. Yet, it has important underlying symmetries and
regularities. Both its unpredictable nature, and its regular hierarchical
structure is a result of the form of the nonlinear interactions. Therefore,
the study of turbulence is a study of the nonlinear interaction, and
the attempt to understand the hierarchical structure of the underlying
symmetries it implies and their limitations\cite{falkovich:09}.

While the simple picture of a turbulent cascade, introduced by Kolmogorov,
involves interactions between different ``scales'' (i.e. wave-number
magnitudes $k$) of a conserved quantity, the Navier-Stokes equation
does not readily uphold this picture. One usually has to write down
the equation for a conserved quadratic quantity, such as energy or
kinetic helicity, and assume statistical isotropy, homogeneity etc.
in order to arrive at a description that is literally consistent with
the basic cascade picture\cite{frisch}. However, it is clear that
the nonlinear cascade happens in the original equation, even without
these assumptions. For instance, even without any assumption of isotropy,
the energy is transferred from wave-number to wave-number. If one
uses a representation of the wave-vector in spherical polar coordinates
in $k$-space {[}i.e. using $k$, $\theta_{k}$, $\phi_{k}$, such
that $\left(k_{x},k_{y},k_{z}\right)$=$\left(k\sin\theta_{k}\cos\phi_{k},k\sin\theta_{k}\sin\phi_{k},k\cos\theta_{k}\right)$
{]}, one could describe how the energy would be transfered from $k$
to $k'$, which is closely related, to what we call the ``cascade'',
even if the cascade as such, is not the only thing that is implied
by the nonlinear interaction.
\begin{figure}
\begin{centering}
\includegraphics[width=8cm]{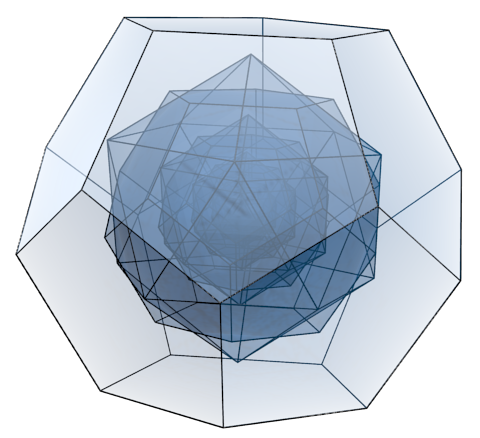}
\par\end{centering}
\caption{\label{fig:alter}Alternating dodecohedron-icosahedron shells covering
the Fourier space. Each $\mathbf{k}$ starts at the origin and ends
at one of the vertices of this object.}
\end{figure}

It is therefore tempting to imagine a discretization of the $k$ space
using some form of spherical polar coordinates, which would assign
the phenomenon of nonlinear cascade to a particular direction $k$.
Furthermore, one may introduce a logarithmic discretization in this
direction, so that with only a small number of points, one may cover
a large range in $k$. In the study of weak wave turbulence, where
the frequency and the wave-number can be linked using a dispersion
relation, such logarithmic grid may be used without any difficulty
(e.g. \cite{galtier:00} for weak MHD turbulence). It is also commonly
used in isotropic cascade modelling using closures, such as eddy damped
quasi-normal Markovian approximation \cite{leith:71,bowman:96,meldi:12},
or differential approximation models \cite{lilly:89,lvov:06,thiagalinam:12}.
Cascade models that use a logaritmic discretization, or shell models
(see for example Ref. \cite{biferale:03}), are also simplified models
that try to exploit this particular aspect of the geometry of the
turbulent cascade. Logarithmically discretized models (LDMs) for two
dimensional turbulence, which can be derived from a systematic self-similar
reduction of the Fourier space\cite{gurcan:16a}, can also be counted
among these models. 

More generally, the reduction of a continuous, but self-similar system,
to a finite number of interacting modes \cite{zeitlin:91}, respecting
the original self-similar structure, has been studied in the past
for cascade models based on the structure of the Burger's equation
\cite{kerr:78} or Navier-Stokes \cite{eggers:91}, especially in
the context of earlier high resolution simulation efforts \cite{vazquez-semadeni:92}.
Such reduction procedures played an important role, beyond simple
numerical convenience, in turbulence studies by providing theoretical
insight into the underlying hierarchical structure of the dynamics
of turbulence \cite{frisch:78,biferale:95,lvov:98}. Since one of
the primary goals of turbulence study, is a reduction of the degrees
of freedom in turbulent dynamics as faithfully as possible to its
essential features, such models were studied for various aspects.
It was found for instance that, severely reduced models, such as shell
models, which rely on a truncation of the original system to keep
only local interactions, recover both the wave-number spectrum and
its intermittency \cite{ohkitani:89,jensen:91,pisarenko:93}. Even
though how the shell model gets the intermittency correction is still
contraversial.

In the same spirit, here we present a direct discretization of the
Navier-Stokes equation in $\mathbf{k}$-space on a special mesh constructed
from self-similarly scaled dodecahedron-icosahedron compounds (i.e.
Wenninger model index 47 \cite{wenninger:book:74}). As will be shown
below, choosing the scaling between two consecutive dodecahedron-icosahedron
compounds equal to the \emph{golden ratio}, allows the possibility
of forming triangles using only discretized wave-vectors. The same
grid can be obtained by considering nested, alternating icosahedra
and dodecahedra where the scaling between the inner dodecahedron and
the outer icosahedron is the square-root of the golden ratio times
the factor $\sqrt{\sqrt{5}/3}$ so that the two consecutive polyhedra
are the the duals of one another.

The model as introduced here may appear artificial as it is composed
of rather complicated sounding polyhedra. However it is a minimal
model in the sense that the icosahedron is the minimal basis for the
\emph{icosphere}, which is an approximation to the sphere with roughly
equal vertex density everwhere on its surface (unlike a straightforward
discretization of angles, which results in higher vertex density near
the poles), and that its dual polyhedron, the dodecahedron is necessary
for completing the triads formed by the position vectors (in $k$-space)
of its vertices.

The method of reduction, which could be generalized, can be thought
of a \emph{discretization based on wave-vector-triads} instead of
a more classical discretization based on wave-vectors themselves.
Notice that, the emposed self similar structure of the model enforces
a uniform triad density as a function of scale, which is known to
be incorrect in the case of turbulence, and appears as an important
weakness of the model. However the model gives the correct wave-number
spectrum and it can describe anisotropy in three dimensions. Furthermore,
a higher order method, based on algorithmic construction of the grid
may be imagined where wave-vectors picked from two distant (i.e. non-consecutive)
polyhedra could be used to deduce a third polyhedron by completing
the triads.

It can be speculated that, such a structure may be meaningful beyond
the model itself as a crystalline state in wave-number space, which
could be tailored by choosing the initial conditions and the driving
to fall on the vertices of a compound polyhedron and the vertices
of such an object would interact with eachother to drive other objects
that are self-similar scalings of the initial object as well as other
higher order compound objects.

\section{The Nested Polyhedra Model}

Consider the Navier-Stokes equation in Fourier space:
\begin{equation}
\partial_{t}u_{\mathbf{k}}^{i}+ik_{\kappa}\left[\delta_{ij}-\frac{k_{i}k_{j}}{k^{2}}\right]\sum_{\mathbf{p}+\mathbf{q}=-\mathbf{k}}u_{\mathbf{p}}^{\kappa*}u_{\mathbf{q}}^{j*}=0\mbox{.}\label{eq:n-s}
\end{equation}
We propose a discretization of the $k$ space using a logarithmic
alternating icosahedral/dodecahedral basis (see figure \ref{fig:alter}):
\[
\mathbf{k}=k_{n}\hat{\mathbf{k}}_{\ell}
\]
where $k_{n}=g^{n}\lambda k_{0}$ is the logarithmically spaced wavenumber
magnitude with $g=\sqrt{\left(1+\sqrt{5}\right)/2}$ ,
\[
\lambda=\begin{cases}
\sqrt{\frac{\sqrt{5}}{3}} & \mbox{for icosahedron}\\
1 & \mbox{for dodecahedron}
\end{cases}
\]
and
\begin{equation}
\hat{\mathbf{k}}_{\ell}=e_{\ell}^{j}=\left[\sin\theta_{\ell}\cos\phi_{\ell},\sin\theta_{\ell}\sin\phi_{\ell},\cos\theta_{\ell}\right]\label{eq:kl}
\end{equation}
where $\theta_{\ell}$ and $\phi_{\ell}$ are to be picked from the
angles corresponding to the icosahedral and the dodecahedral vertices,
listed in table \ref{tab:angles}. It is shown below that this choice
comes from the condition of forming triads with the vertices of the
three consecutive polyhedra. 
\begin{table}
\begin{tabular}{ccccccccccccccc}
$\ell$ &  & $\theta_{\ell}$ & $\phi_{\ell}$ & \hspace{0.4cm} &  &  &  &  &  & \hspace{1cm} & $\ell$ &  & $\theta_{\ell}$ & $\phi_{\ell}$\tabularnewline
\cline{3-4} \cline{14-15} 
0 &  & $\alpha$ & $\pi/5$ &  & 10 &  & $\pi-\alpha$ & $6\pi/5$ &  &  & 0 &  & $0$ & $.$\tabularnewline
\cline{14-15} 
1 &  & $\alpha$ & $3\pi/5$ &  & 11 &  & $\pi-\alpha$ & $8\pi/5$ &  &  & 1 &  & $\gamma$ & $0$\tabularnewline
2 &  & $\alpha$ & $\pi$ &  & 12 &  & $\pi-\alpha$ & 0 &  &  & 2 &  & $\gamma$ & $2\pi/5$\tabularnewline
3 &  & $\alpha$ & $7\pi/5$ &  & 13 &  & $\pi-\alpha$ & $2\pi/5$ &  &  & 3 &  & $\gamma$ & $4\pi/5$\tabularnewline
4 &  & $\alpha$ & $9\pi/5$ &  & 14 &  & $\pi-\alpha$ & $4\pi/5$ &  &  & 4 &  & $\gamma$ & $6\pi/5$\tabularnewline
\cline{3-4} \cline{8-9} 
5 &  & $\beta$ & $\pi/5$ &  & 15 &  & $\pi-\beta$ & $6\pi/5$ &  &  & 5 &  & $\gamma$ & $8\pi/5$\tabularnewline
\cline{14-15} 
6 &  & $\beta$ & $3\pi/5$ &  & 16 &  & $\pi-\beta$ & $8\pi/5$ &  &  & 6 &  & $\pi$ & .\tabularnewline
\cline{14-15} 
7 &  & $\beta$ & $\pi$ &  & 17 &  & $\pi-\beta$ & 0 &  &  & 7 &  & $\pi-\gamma$ & $\pi$\tabularnewline
8 &  & $\beta$ & $7\pi/5$ &  & 18 &  & $\pi-\beta$ & $2\pi/5$ &  &  & 8 &  & $\pi-\gamma$ & $7\pi/5$\tabularnewline
9 &  & $\beta$ & $9\pi/5$ &  & 19 &  & $\pi-\beta$ & $4\pi/5$ &  &  & 9 &  & $\pi-\gamma$ & $9\pi/5$\tabularnewline
 &  &  &  &  &  &  &  &  &  &  & 10 &  & $\pi-\gamma$ & $\pi/5$\tabularnewline
 &  &  &  &  &  &  &  &  &  &  & 11 &  & $\pi-\gamma$ & $3\pi/5$\tabularnewline
\end{tabular}

\caption{\label{tab:angles}Polar and azimuthal angles $\theta$ and $\phi$
of a dodecahedron (left) and an icosahedron (right) are listed. Here
$\alpha=\arcsin\left(\varphi/\sqrt{3}\right)-\arccos\left(\varphi/\sqrt{\varphi+2}\right)$
and $\beta=\arctan\left(2\varphi^{2}\right)$ and $\gamma=\pi/2-\arctan\left(1/2\right)$
with $\varphi=\left(1+\sqrt{5}\right)/2$. }
\end{table}
\begin{table}
\vspace{0.5cm}
\begin{tabular}{ccccccc}
$\ell_{i}$:$\left(n\right)$ & $\ell_{d}^{'}$:$\left(n-1\right)$ & $\ell_{d}^{''}$:$\left(n+1\right)$ & \hspace{0.5cm} & $\ell_{i}$:$\left(n\right)$ & $\ell_{d}^{'}$:$\left(n-1\right)$ & $\ell_{d}^{''}$:$\left(n+1\right)$\tabularnewline
\cmidrule{1-3} \cmidrule{5-7} 
0 & 5 & 10 &  & 3 & 0 & 11\tabularnewline
 & 6 & 11 &  &  & 18 & 16\tabularnewline
 & 7 & 12 &  &  & 14 & 9\tabularnewline
 & 8 & 13 &  &  & 15 & 17\tabularnewline
 & 9 & 14 &  &  & 3 & 12\tabularnewline
\cmidrule{1-3} \cmidrule{5-7} 
1 & 1 & 10 &  & 4 & 1 & 12\tabularnewline
 & 18 & 15 &  &  & 19 & 17\tabularnewline
 & 12 & 7 &  &  & 10 & 5\tabularnewline
 & 16 & 19 &  &  & 16 & 18\tabularnewline
 & 3 & 14 &  &  & 4 & 13\tabularnewline
\cmidrule{1-3} \cmidrule{5-7} 
2 & 4 & 10 &  & 5 & 2 & 13\tabularnewline
 & 17 & 15 &  &  & 15 & 18\tabularnewline
 & 13 & 8 &  &  & 11 & 6\tabularnewline
 & 19 & 16 &  &  & 17 & 19\tabularnewline
 & 2 & 11 &  &  & 0 & 14\tabularnewline
\cmidrule{1-3} \cmidrule{5-7} 
\end{tabular}

\caption{\label{tab:idd}Two dodecahedral vertices, from the neighbouring shells
(i.e. $n-1$ and $n+1$) that form a perfect triad with the icosahedral
vertex at shell $n$. }
\end{table}

Note that by enumerating the vertices of the icosahedron and the dodecahedron,
we reduce the number of indices necessary to describe a given wave-vector
from $3$ to $2$ (i.e. using only $n$ and $\ell$, we can define
a unique wave-vector). It is also important to mention that since
the original velocity field $\mathbf{u}\left(\mathbf{x},t\right)$
is real, its Fourier transform has the symmetry that $\mathbf{u}\left(\mathbf{k},t\right)=\mathbf{u}^{*}\left(-\mathbf{k},t\right)$.
The assignment of numbers to vertices are picked such that $\hat{\mathbf{k}}_{\ell+N_{\ell}/2}=-\hat{\mathbf{k}}_{\ell}$
where $N_{\ell}$ is the number of vertices of the polyhedron in consideration
(i.e. $N_{\ell}=12$ for the icosahedron, while $N_{\ell}$=20 for
the dodecahedron). This symmetry can be used to reduce the number
of degrees of freedom by half. Otherwise, one must pay attention that
the initial conditions as well as all the terms in the equation (such
as forcing, dissipation etc.) respect this symmetry.

Three dimensional turbulence requires solving three vector components
of the velocity. Here we use cartesian coordinates $\mathbf{u}_{\mathbf{k}}=u_{k_{n}\hat{\mathbf{k}}_{\ell}}^{\left(x\right)}\hat{\mathbf{x}}+u_{k_{n}\hat{\mathbf{k}}_{\ell}}^{\left(y\right)}\hat{\mathbf{y}}+u_{k_{n}\hat{\mathbf{k}}_{\ell}}^{\left(z\right)}\hat{\mathbf{z}}\rightarrow u_{n\ell}^{i}\hat{\mathbf{x}}_{i}$.
In this representation, the Navier-Stokes equation becomes:
\begin{equation}
\partial_{t}u_{n,\ell}^{i}+ik_{n\ell}^{\kappa}\left[\delta_{ij}-\frac{k_{n\ell}^{i}k_{n\ell}^{j}}{k_{n}^{2}}\right]\sum_{n',\ell'}u_{n'\ell'}^{\kappa*}u_{n''\ell''}^{j*}=0\label{eq:ns}
\end{equation}
where $\ell''$ and $n''$ can be inferred from $n$,$\ell$, $n'$
and $\ell'$ using the fact that the corresponding wavenumbers form
a triad:
\[
k_{n}\hat{\mathbf{k}}_{\ell}+k_{n'}\hat{\mathbf{k}}_{\ell'}+k_{n''}\hat{\mathbf{k}}_{\ell''}=0
\]
consider three consecutive spherical shells such that $n'=n-1$ and
$n''=n+1$, and take alternating spheres to be discretized as dodecahedrons
and icosahedrons (i.e. $n=1$ is an icosahedron, $n=2$ is a dodecahedron,
$n=3$ is icosahedron and so on), we can write:
\begin{equation}
k_{n}\hat{\mathbf{k}}_{\ell}^{i}+k_{n-1}\hat{\mathbf{k}}_{\ell'}^{d}+k_{n+1}\hat{\mathbf{k}}_{\ell''}^{d}=0\label{eq:cond1}
\end{equation}
\begin{equation}
k_{n}\hat{\mathbf{k}}_{\ell}^{d}+k_{n-1}\hat{\mathbf{k}}_{\ell'}^{i}+k_{n+1}\hat{\mathbf{k}}_{\ell''}^{i}=0\label{eq:cond2}
\end{equation}
which can be shifted in $n$ (i.e. $n\rightarrow n+1$ and $n\rightarrow n-1$)
to cover all the necessary triads. Note that other interactions don't
correspond to grid points and will be dropped. It is interesting to
note that this does not lead to leaking of conserved quantites.

Now consider $\ell=0$, $\ell'=5$, $\ell''=10$ for the first equation:
\begin{align*}
 & \left[g\sin\left(\pi-\alpha\right)\cos\frac{6\pi}{5}+g^{-1}\sin\beta\cos\frac{\pi}{5}\right]\hat{\mathbf{x}}\\
 & +\left[g\sin\left(\pi-\alpha\right)\sin\frac{6\pi}{5}+g^{-1}\sin\beta\sin\frac{\pi}{5}\right]\hat{\mathbf{y}}\\
 & +\left[\lambda+g^{-1}\cos\beta+g\cos\left(\pi-\alpha\right)\right]\hat{\mathbf{z}}=0
\end{align*}
Where the coefficients of $\hat{\mathbf{x}}$ and $\hat{\mathbf{y}}$
can be made to vanish by choosing $g=\sqrt{\varphi}=\sqrt{\left(1+\sqrt{5}\right)/2}$,
while in order to make the coefficient of $\hat{\mathbf{z}}$ vanish,
we need to choose the scaling of the radius of the icosahedron with
respect to $g^{n}k_{0}$ as  $\lambda=\sqrt{\frac{\sqrt{5}}{3}}$.
This way we can satisfy the condition of the triad. The icosahedron
that is constructed this way is actually nothing but the dual icosahedron
of the inner dodecahedron with the radius $k_{n-1}$. While these
two can be thought of as being on separate shells in $k$-space, since
their radii are different. They could also be thought of sampling
a single shell together in the form of a dodecahedron-icosahedron
compound (the shell boundary in this case could be thought to be in
between the two consecutive compounds).
\begin{table}
\vspace{0.5cm}
\begin{tabular}{ccccccc}
$\ell_{d}$:$\left(n\right)$ & $\ell_{i}^{'}$:$\left(n-1\right)$ & $\ell_{i}^{''}$:$\left(n+1\right)$ & \hspace{0.5cm} & $\ell_{d}$:$\left(n\right)$ & $\ell_{i}^{'}$:$\left(n-1\right)$ & $\ell_{i}^{''}$:$\left(n+1\right)$\tabularnewline
\cmidrule{1-3} \cmidrule{5-7} 
0 & 4 & 6 &  & 5 & 3 & 8\tabularnewline
 & 9 & 7 &  &  & 5 & 7\tabularnewline
 & 11 & 8 &  &  & 6 & 4\tabularnewline
\cmidrule{1-3} \cmidrule{5-7} 
1 & 5 & 6 &  & 6 & 1 & 8\tabularnewline
 & 7 & 9 &  &  & 4 & 9\tabularnewline
 & 10 & 8 &  &  & 6 & 5\tabularnewline
\cmidrule{1-3} \cmidrule{5-7} 
2 & 1 & 6 &  & 7 & 2 & 9\tabularnewline
 & 11 & 9 &  &  & 5 & 10\tabularnewline
 & 8 & 10 &  &  & 6 & 1\tabularnewline
\cmidrule{1-3} \cmidrule{5-7} 
3 & 2 & 6 &  & 8 & 3 & 10\tabularnewline
 & 9 & 11 &  &  & 1 & 11\tabularnewline
 & 7 & 10 &  &  & 6 & 2\tabularnewline
\cmidrule{1-3} \cmidrule{5-7} 
4 & 3 & 6 &  & 9 & 2 & 7\tabularnewline
 & 8 & 11 &  &  & 4 & 11\tabularnewline
 & 10 & 7 &  &  & 6 & 3\tabularnewline
\cmidrule{1-3} \cmidrule{5-7} 
\end{tabular}

\caption{\label{tab:dii}Two icosahedral vertices, from the neighbouring shells
(i.e. $n-1$ and $n+1$) that form a perfect triad with the dodecahedral
vertex at shell $n$.  }
\end{table}

Of course one can rotate the triangle around the primary vector $\mathbf{k}$
to obtain another interacting pair. For instance for the node $\ell=0$,
the condition of the triad will be satisfied by the pairs $\left\{ \ell',\ell''\right\} =\left[\left\{ 5,10\right\} ,\left\{ 6,11\right\} ,\left\{ 7,12\right\} ,\left\{ 8,13\right\} ,\left\{ 9,14\right\} \right]$.
Since each point of the icosahedron is equivalent, we can compute
the interacting pairs of dodecahedral vertices for each vertex of
the icosahedron using the same algorithm. The node-pair connections
obtained this way are given in tables \ref{tab:idd} and \ref{tab:dii}.

Now consider (\ref{eq:cond2}), and choose $\ell=0$, $\ell'=6$,
$\ell''=4$. This gives:
\begin{align*}
 & \left(\sin\alpha\cos\frac{\pi}{5}+g^{-1}\lambda\sin\gamma\cos\frac{6\pi}{5}\right)\hat{\mathbf{x}}\\
 & +\left(\sin\alpha\sin\frac{\pi}{5}+g^{-1}\lambda\sin\gamma\sin\frac{6\pi}{5}\right)\hat{\mathbf{y}}\\
 & +\left(\cos\alpha+g^{-1}\lambda\cos\gamma-\lambda g\right)\hat{\mathbf{z}}=0
\end{align*}
which is automatically satisfied by the earlier choice $g=\sqrt{\varphi}$
and $\lambda=\sqrt{\sqrt{5}/3}$. Note again that the above condition
for $\ell=0$ is also satisfied by the pairs: $\left\{ \ell',\ell''\right\} =\left[\left\{ 4,6\right\} ,\left\{ 9,7\right\} ,\left\{ 11,8\right\} \right]$.
Similarly as the case where the icosahedron was in the middle, we
can find the rest of the interacting pairs of icosahedral vertices
for each vertex of the dodecahedron by rotating the mesh.
\begin{figure}
\begin{centering}
\includegraphics[width=4cm]{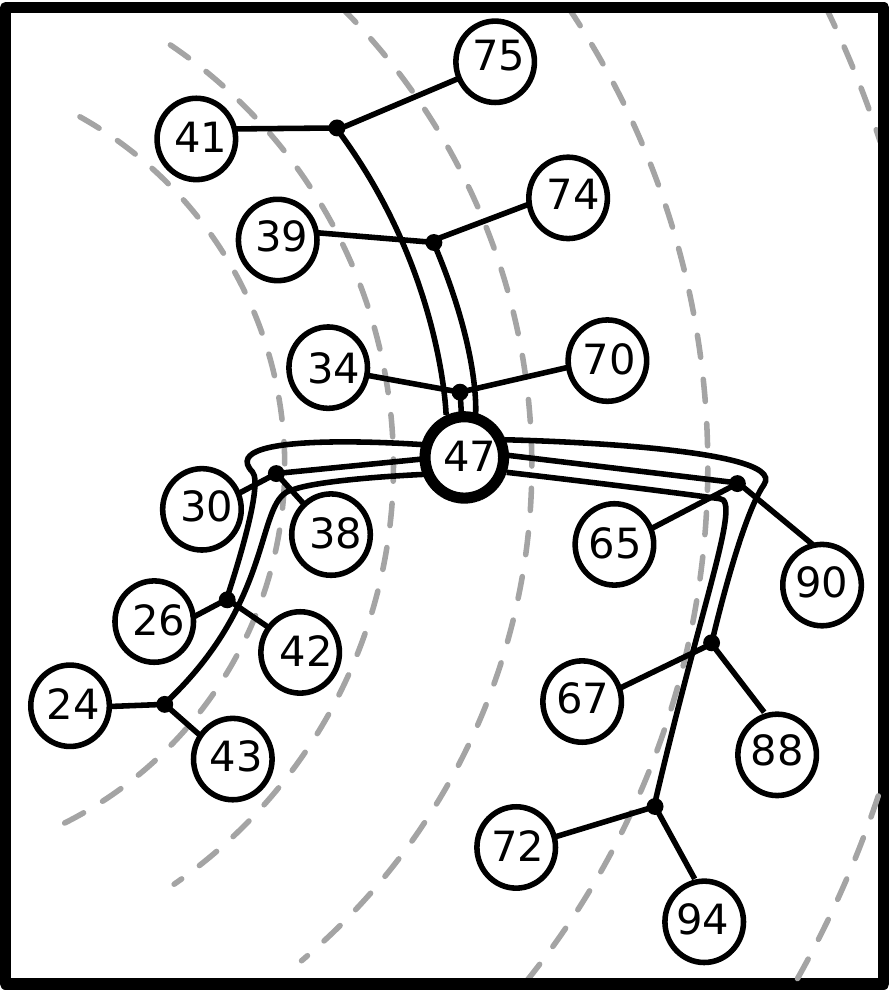}
\par\end{centering}
\caption{\label{fig:network}Pairs of nodes interacting with the node number
$47$. (i.e. node $\ell=3$ on the shell $n=3$) where the first shell
is an icosahedron, which is shown here as an example. }
\end{figure}
\begin{figure}
\begin{centering}
\includegraphics[width=0.99\columnwidth]{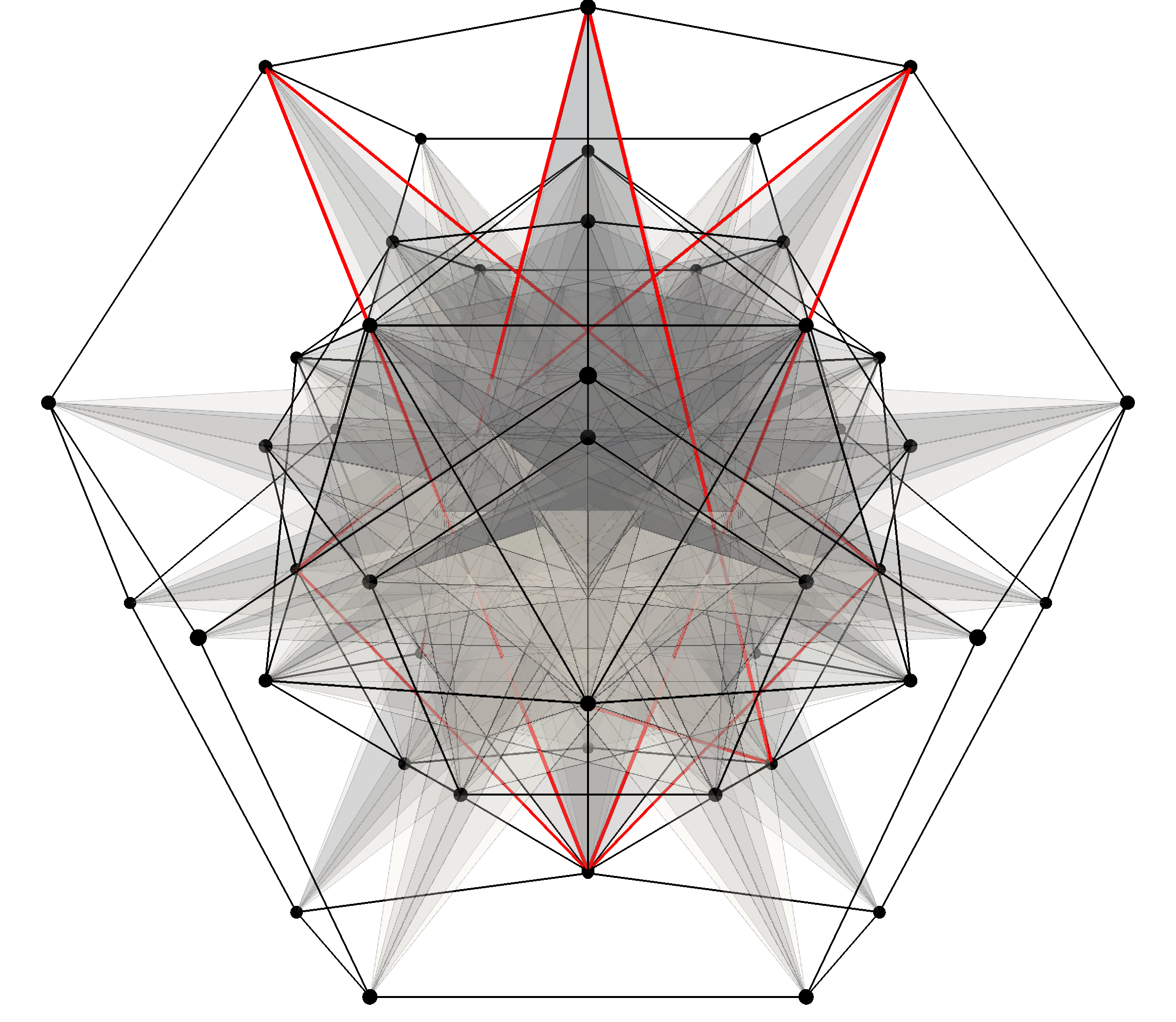}
\par\end{centering}
\caption{All the k-space triads corresponding to an icosahedron squeezed in
between two dodecahedra. }
\end{figure}
\begin{figure}
\begin{centering}
\includegraphics[width=0.99\columnwidth]{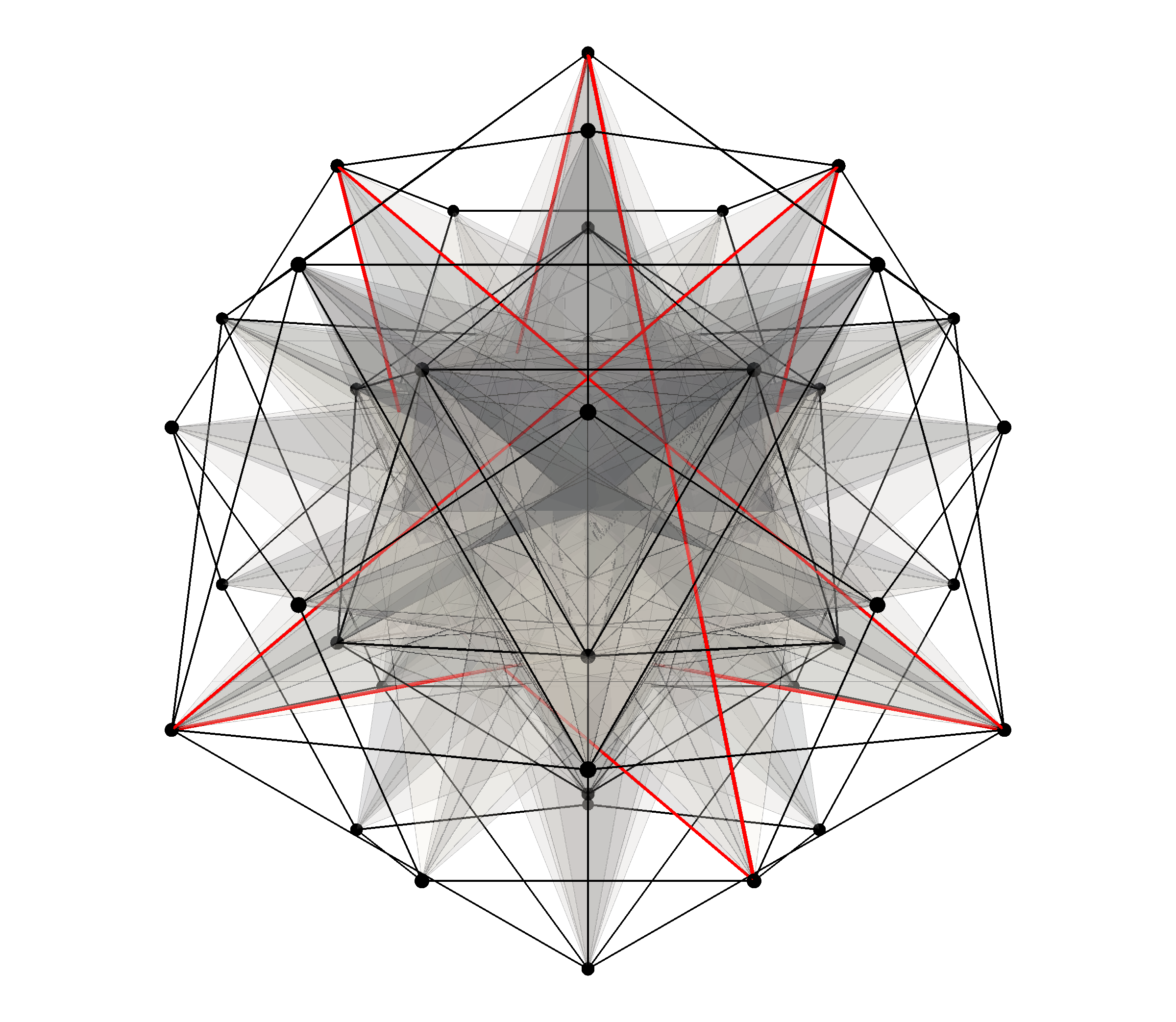}
\par\end{centering}
\caption{All the k-space triads corresponding to a dodecahedron squeezed in
between two icosahedra. }
\end{figure}

Since exchanging $\left(n',\ell'\right)\leftrightarrow\left(n'',\ell''\right)$,
another interaction is obtained, we will consider this explicitly
by symmetrizing the equations as:
\begin{equation}
\partial_{t}u_{n\ell}^{i}+iM_{n\ell}^{\kappa ij}\sum_{n'<n'',\ell'}\left(u_{n'\ell'}^{\kappa*}u_{n''\ell''}^{j*}+u_{n''\ell''}^{\kappa*}u_{n'\ell'}^{j*}\right)=0\label{eq:nssym}
\end{equation}
where
\[
M_{n\ell}^{\kappa ij}=k_{n\ell}^{\kappa}\left[\delta_{ij}-\frac{k_{n\ell}^{i}k_{n\ell}^{j}}{k_{n}^{2}}\right]
\]

This way we can go over each node-pair connection once, without paying
attention to the sign, and all possible interactions will be covered.
Defining $n$ as the flattened node number (e.g. $\left\{ n,\ell\right\} =\left\{ 3,3\right\} \rightarrow n=12\times2+20+3=47$,
if the first shell is an icosahedron), instead of the shell number
as before.
\begin{equation}
\partial_{t}u_{n}^{i}+iM_{n}^{\kappa ij}\sum_{\left\{ n',n''\right\} =\mathbf{p}_{n}}\left(u_{n'}^{\kappa*}u_{n''}^{j*}+u_{n''}^{\kappa*}u_{n'}^{j*}\right)=0\;\mbox{.}\label{eq:nsnw}
\end{equation}
where the sum is computed over the interacting $\left\{ n',n''\right\} $
pairs of a node $n$ (e.g. $\mathbf{p}_{47}$ is the list of the pairs
of nodes shown in figure \ref{fig:network}). These connections can
be obtained using the tables \ref{tab:idd} and \ref{tab:dii} and
the flattening rule $m=\mbox{floor}\left(n/2\right)\times32+\left(n\mod2\right)\times N_{fs}+\ell$
(and then $m\rightarrow n$), where $N_{fs}$ is the number of vertices
of the first shell, and can be thought of as a regular network model
(see figure \ref{fig:network}). Note that, in (\ref{eq:nsnw}) the
interaction matrix $M_{n\ell}^{\kappa ij}\rightarrow M_{n}^{\kappa ij}$
is also flattened the same way as the vector $u_{n\ell}^{i}\rightarrow u_{n}^{i}$
.

\subsection{Further simplifications\label{subsec:Further-simplifications}}

As discussed earlier, in order to reduce the degrees of freedom of
the nested polyhedra model of turbulence, one may consider only half
of each polyhedra (for instance the $k_{z}>0$ hemisphere), and obtain
the other half using the relation $\mathbf{u}_{-\mathbf{k}}=\mathbf{u}_{\mathbf{k}}^{*}$.
In order to achieve this, in practice one has to keep in mind wheter
a node in an interaction is conjugated or not. (i.e. in tables \ref{tab:idd}
and \ref{tab:dii} if the node number falls into the upper hemisphere
it would be conjugated, if it falls into the lower hemisphere it would
not be conjugated in eqn. \ref{eq:nsnw}). This can be done by keeping
a ``conjugated flag'' for each interaction between nodes. In addition
to decreasing the number of degrees of freedom to half, this approach
has the advantage of automatically imposing the reality condition
of the velocity field as a function of space, which is important and
is not exact in the more straightforward formulation.

Another more interesting simplification is to consider the helical
decomposition, which allows the reduction of the velocity 3-vector
to two scalars representing right handed and left handed helicities
with respect to the wavenumber. This is possible because of the fact
that the velocity field is divergence free in the Navier-Stokes equation,
and therefore its projection onto the wavenumber has to vanish. This
formulation allows us to reduce the number of degrees of freedom to
2/3 times the initial number and it guarantees that the velocity field
remains divergence free, in contrast to the straightforward method,
which does not guarantee this numerically.

Together these two simplifications would permit a reduction of the
number of degrees of freedom to 1/3 times the initial number, and
guarantee a real and divergence free velocity field as function of
space. 

Note that a simple spherical representation of the velocity field
in $\mathbf{k}$ space (i.e. $\mathbf{u}_{n\ell}=u_{n\ell}^{k}\hat{\mathbf{k}}_{n\ell}+u_{n\ell}^{\theta}\hat{\boldsymbol{\theta}}_{n\ell}+u_{n\ell}^{\phi}\hat{\boldsymbol{\phi}}_{n\ell}$)
also reduces the velocity field to two dimensions since the ``radial''
component vanihses (i.e. $u_{n\ell}^{k}=\mathbf{u}_{n\ell}\cdot\hat{\mathbf{k}}_{n\ell}=0$).
However the simplification of the interaction coefficients as well
as the direct physical interpretation in terms of helicity are lost
in this representation. Nonetheless, this representation allows us
to see that the Fourier transformed velocity field is everywhere tangential
to the spherical shells in k-space.

Further reduction can be achieved by associating one kind of helicity
(i.e. Right), with one type of polyhedron (i.e. dodecahedron) and
the other kind of helicity (i.e. Left) with the other type of polyhedron
(i.e. icosahedron). This reduces the number of degrees of freedom,
further by half. It also assigns a physical sense to the different
types of polyhedra in the model. The reduced model obtained this way
can be written as:
\begin{align*}
\partial_{t}u_{n\ell}^{s_{n}}+ & \frac{1}{4}\sum_{n'<n'',\ell'}\left(k_{n'}s_{n'}-k_{n''}s_{n''}\right)\\
 & \times\left(\hat{\mathbf{h}}_{\ell}^{s_{n}*}\cdot\hat{\mathbf{h}}_{\ell'}^{s_{n'}*}\times\hat{\mathbf{h}}_{\ell''}^{s_{n''}*}\right)u_{n'\ell'}^{s_{n'}*}u_{n''\ell''}^{s_{n''}*}=-\nu k_{n}^{2}u_{n,\ell}^{s_{n}}
\end{align*}

where 
\begin{equation}
\mathbf{u}\left(\mathbf{x}\right)=\sum_{n=0}^{N}u_{n\ell}^{s_{n}}\hat{\mathbf{h}}_{\ell}^{s_{n}}\label{eq:wal1}
\end{equation}
and $s_{n}=\left(+\right)$ if $n$:even, and $s_{n}=\left(-\right)$
if $n$:odd, with:
\[
\hat{\mathbf{h}}_{n\ell}^{\pm}=\hat{\boldsymbol{\nu}}_{\mathbf{k}}\times\hat{\mathbf{k}}\pm i\hat{\boldsymbol{\nu}}_{\mathbf{k}}
\]
\[
\hat{\boldsymbol{\nu}}_{\mathbf{k}}=\frac{\mathbf{k}\times\hat{\mathbf{z}}}{\left|\mathbf{k}\times\hat{\mathbf{z}}\right|}
\]
 This gives\begin{widetext}
\begin{align*}
\partial_{t}u_{n,\ell}^{\left(+\right)} & +k_{n}g^{-2}\left(1+\lambda g\right)\sum_{\left\{ \ell',\ell''\right\} }\left(\hat{\mathbf{h}}_{\ell}^{\left(+\right)*}\cdot\hat{\mathbf{h}}_{\ell'}^{\left(+\right)*}\times\hat{\mathbf{h}}_{\ell''}^{\left(-\right)*}\right)u_{n-2,\ell'}^{\left(+\right)*}u_{n-1,\ell''}^{\left(-\right)*}\\
 & -\lambda k_{n}g^{-1}\left(g^{2}-1\right)\sum_{\left\{ \ell',\ell''\right\} }\left(\hat{\mathbf{h}}_{\ell}^{\left(+\right)*}\cdot\hat{\mathbf{h}}_{\ell'}^{\left(-\right)*}\times\hat{\mathbf{h}}_{\ell''}^{\left(-\right)*}\right)u_{n-1,\ell'}^{\left(-\right)*}u_{n+1,\ell''}^{\left(-\right)*}\\
 & -k_{n}g\left(\lambda+g\right)\sum_{\left\{ \ell',\ell''\right\} }\left(\hat{\mathbf{h}}_{\ell}^{\left(+\right)*}\cdot\hat{\mathbf{h}}_{\ell'}^{\left(-\right)*}\times\hat{\mathbf{h}}_{\ell''}^{\left(+\right)*}\right)u_{n+1,\ell'}^{\left(-\right)*}u_{n+2,\ell''}^{\left(+\right)*}\\
 & =-\nu k_{n}^{2}a_{n,\ell}^{\left(+\right)}
\end{align*}
for even $n$, and 
\begin{align*}
\partial_{t}u_{n,\ell}^{\left(-\right)} & -k_{n}g^{-2}\left(\lambda+g\right)\sum_{\left\{ \ell',\ell''\right\} }\left(\hat{\mathbf{h}}_{\ell}^{\left(-\right)*}\cdot\hat{\mathbf{h}}_{\ell'}^{\left(-\right)*}\times\hat{\mathbf{h}}_{\ell''}^{\left(+\right)*}\right)u_{n-2,\ell'}^{\left(-\right)*}u_{n-1,\ell''}^{\left(+\right)*}\\
 & +k_{n}g^{-1}\left(1-g^{2}\right)\sum_{\left\{ \ell',\ell''\right\} }\left(\hat{\mathbf{h}}_{\ell}^{\left(-\right)*}\cdot\hat{\mathbf{h}}_{\ell'}^{\left(+\right)*}\times\hat{\mathbf{h}}_{\ell''}^{\left(+\right)*}\right)u_{n-1,\ell'}^{\left(+\right)*}u_{n+1,\ell''}^{\left(+\right)*}\\
 & +k_{n}g\left(1+\lambda g\right)\sum_{\left\{ \ell',\ell''\right\} }\left(\hat{\mathbf{h}}_{\ell}^{\left(-\right)*}\cdot\hat{\mathbf{h}}_{\ell'}^{\left(+\right)*}\times\hat{\mathbf{h}}_{\ell''}^{\left(-\right)*}\right)u_{n+1,\ell'}^{\left(+\right)*}u_{n+2,\ell''}^{\left(-\right)*}\\
 & =-\nu k_{n}^{2}u_{n,\ell}^{\left(-\right)}
\end{align*}
for odd $n$. \end{widetext}It is interesting to note that these
two equations have the same form (apart from the additional $\ell$
resolution) as the model discussed in Ref. \cite{depietro:15}. In
particular the above form corresponds to the model SM1 as discussed
in that paper. 

\subsection{Conservation Laws}

Energy conservation can be shown by considering the energy of a single
triad (e.g. $\mathbf{k}_{n\ell}$, \textbf{$\mathbf{k}_{n-1,\ell'}$}
and $\mathbf{k}_{n+1,\ell''}$):
\begin{align*}
\frac{dE_{\Delta}}{dt}= & \mbox{Re}\bigg[i\overline{M}_{n\ell}^{\kappa ij}u_{n-1,\ell'}^{\kappa*}u_{n+1,\ell''}^{j*}u_{n\ell}^{i*}\\
 & +i\overline{M}_{n-1,\ell'}^{\kappa ij}u_{n-1,\ell'}^{i*}u_{n,\ell}^{\kappa*}u_{n+1,\ell''}^{j*}\\
 & +i\overline{M}_{n+1,\ell''}^{\kappa ij}u_{n+1,\ell''}^{i*}u_{n-1,\ell'}^{\kappa*}u_{n,\ell}^{j*}\bigg]=0
\end{align*}
by using the form of $\overline{M}_{n\ell}^{\kappa ij}$ and the facts
that $k_{n\ell}^{i}u_{n\ell}^{i}=0$ and $k_{n\ell}^{i}=-k_{n-1,\ell'}^{i}-k_{n+1,\ell''}^{i}$.
The total energy can then be written as a sum of the energy $E_{\Delta}$
over triads. Since each closed triad conserves energy, any discretization
of the Fourier space using a reduced set of ``triads'' automatically
respects energy conservation. In fact this should be true for all
the conservation laws of the system even without an explicit knowledge
of the conservation laws. This is important, as we will see consecutively,
since it provides a ``derivation'' of shell models and their generalizations
without a detailed knowledge of the conservation laws. A similar effort
was discussed earliler for 2D turbulence \cite{gurcan:16a}.

\subsection{Connection to Shell Models}

When the sum over different $n$ values are written explicitly, the
model takes the form: 
\begin{align}
\partial_{t}u_{n,\ell}^{i}+i\overline{M}_{n\ell}^{\kappa ij}\sum_{\left\{ \ell',\ell''\right\} }\bigg[ & u_{n-2,\ell'}^{\kappa*}u_{n-1,\ell''}^{j*}+u_{n-1,\ell'}^{\kappa*}u_{n+1,\ell''}^{j*}\nonumber \\
 & +u_{n+1,\ell'}^{\kappa*}u_{n+2,\ell''}^{j*}\bigg]\label{eq:goy_form}
\end{align}
regardless of wheter the $n$th shell is an icosahedron or a dodecahedron.
Note that $\overline{M}_{n\ell}^{\kappa ij}=M_{n\ell}^{\kappa ij}+M_{n\ell}^{ji\kappa}$
and sum is computed over pairs of interacting nodes of the consecutive
shells as given in tables \ref{tab:idd} and \ref{tab:dii}. The basic
form of the equation (\ref{eq:goy_form}) is consistent with the Gledzer-Ohkitani-Yamada
(GOY) model\cite{ohkitani:89}. In fact one can arrive at a form very
similar to the GOY model by arranging a certain (rather particular)
choice of phases and signs of helicities for each node. Taking
\[
u_{n,\ell}^{j}=u_{n}e^{i\theta_{n,\ell}^{j}}
\]
and imposing the resulting coefficients to be independent of $\ell$
(see below for a discussion of this), we get:
\begin{equation}
\partial_{t}u_{n}+ik_{n}\left(a_{n}u_{n-2}^{*}u_{n-1}^{*}+b_{n}u_{n-1}^{*}u_{n+1}^{*}+c_{n}u_{n+1}^{*}u_{n+2}^{*}\right)\label{eq:goy}
\end{equation}
where
\begin{align*}
a_{n} & \equiv\sum_{\left\{ \ell',\ell''\right\} }\hat{\mathbf{k}}_{n\ell}^{j}\left(\delta_{i\kappa}-\hat{\mathbf{k}}_{n\ell}^{\kappa}\hat{\mathbf{k}}_{n\ell}^{i}\right)\left[e^{-i\xi_{n,\ell^{'}\ell^{''}\ell}^{\kappa ji}}+e^{-i\xi_{n,\ell^{'}\ell^{''}\ell}^{j\kappa i}}\right]\\
b_{n} & \equiv\sum_{\left\{ \ell',\ell''\right\} }\hat{\mathbf{k}}_{n\ell}^{j}\left(\delta_{i\kappa}-\hat{\mathbf{k}}_{n\ell}^{\kappa}\hat{\mathbf{k}}_{n\ell}^{i}\right)\left[e^{-i\xi_{n+1,\ell^{'}\ell,\ell^{''}}^{\kappa ij}}+e^{-i\xi_{n+1,\ell^{''}\ell\ell^{'}}^{ji\kappa}}\right]\\
c_{n} & \equiv\sum_{\left\{ \ell',\ell''\right\} }\hat{\mathbf{k}}_{n\ell}^{j}\left(\delta_{i\kappa}-\hat{\mathbf{k}}_{n\ell}^{\kappa}\hat{\mathbf{k}}_{n\ell}^{i}\right)\left[e^{-i\xi_{n+2,\ell\ell^{'}\ell^{''}}^{i\kappa j}}+e^{-i\xi_{n+2,\ell\ell^{''}\ell^{'}}^{ij\kappa}}\right]
\end{align*}
and
\[
\xi_{n,\ell^{'}\ell^{''}\ell}^{\kappa ji}\equiv\theta_{n-2,\ell^{'}}^{\kappa}+\theta_{n-1,\ell^{''}}^{j}+\theta_{n,\ell}^{i}
\]
The fact that (\ref{eq:goy}) loses its dependence on $\ell$ (as
the system is summed over $\ell'$ and $\ell''$) means that the $a_{n}$,
$b_{n}$ and $c_{n}$ as defined above should be identical for all
$\ell$ of a polyhedron. 

Assuming that the phases repeat after each $4$ shells (i.e. $\theta_{n+4,\ell}^{i}=\theta_{n,\ell}^{i}$),
gives $(6+10)\times2=32$ independent nodes and having $2$ independent
vector components for velocity each, we have $64$ independent phases.
The idea that the coefficients of (\ref{eq:goy}) be independent of
$\ell$ can be written as:
\[
a_{n\ell}=a_{n}\;\text{,}\quad b_{n\ell}=b_{n}\;\text{,}\quad c_{n\ell}=c_{n}
\]
for any $\ell$. In total this would give $3\times32=96$ equations
(here $3$ is the number of coefficients per node and $32$ is the
number of nodes. Note that number of components does not enter, since
the coefficients $a_{n\ell}$, $b_{n\ell}$ and $c_{n\ell}$ are already
summed over $i$, $j$ and $\kappa$ ). However the equations for
$c_{n+2,\ell}$, $c_{n+3,\ell}$ and $b_{n+3,\ell}$ involves the
phases $\xi_{n+4,\ell,\ell',\ell''}$, $\xi_{n+5,\ell',\ell,\ell''}$
and $\xi_{n+4,\ell,\ell',\ell''}$ , which are the same as $\xi_{n,\ell,\ell',\ell''}$,
$\xi_{n+1,\ell',\ell,\ell''}$ and $\xi_{n,\ell,\ell',\ell''}$ respectively
(due to the assumption of 4-fold periodicity), resulting in some of
the same equations as before. If we assume that the $n$the shell
is an icosahedron, the number of repeated equations are $\left(2\times10+1\times6\right)\times2=52$
which reduces the number of independent equations to $44$. If the
$n$th shell is a dodecahedron on the other hand, the number of repeated
equations are $\left(2\times6+1\times10\right)\times2=44$, which
results in $52$ independent equations. This means that we can actually
pick the $64$ independent phases in such a way that the $44$ or
$52$ independent equations that guarantees a GOY-like model, are
satisfied, and we would still have $20$ or $12$ undetermined phases,
which could be taken for example as the phases of the independent
components of $n+1$st polyhedron (i.e. $\theta_{n+1,\ell}^{i}$).

Notice that this analytical excercise should qualify as an actual,
rigorous derivation of the GOY model, starting from the nested polyhedra
model, assuming isotropy and imposing some particular phase relations.
Since the nested-polyhedra model comes from a systematic reduction
of the triads in a self similar way, the derivation provides a rigorous
path from the initial field equations to the shell model. Of course
both the coefficients and the shell spacing are not free parameters
as they are imposed by the constraints of self similarity of the nested
polyhedra model.

\subsection{Velocity field as function of space}

\begin{figure}
\begin{centering}
\includegraphics[width=8cm]{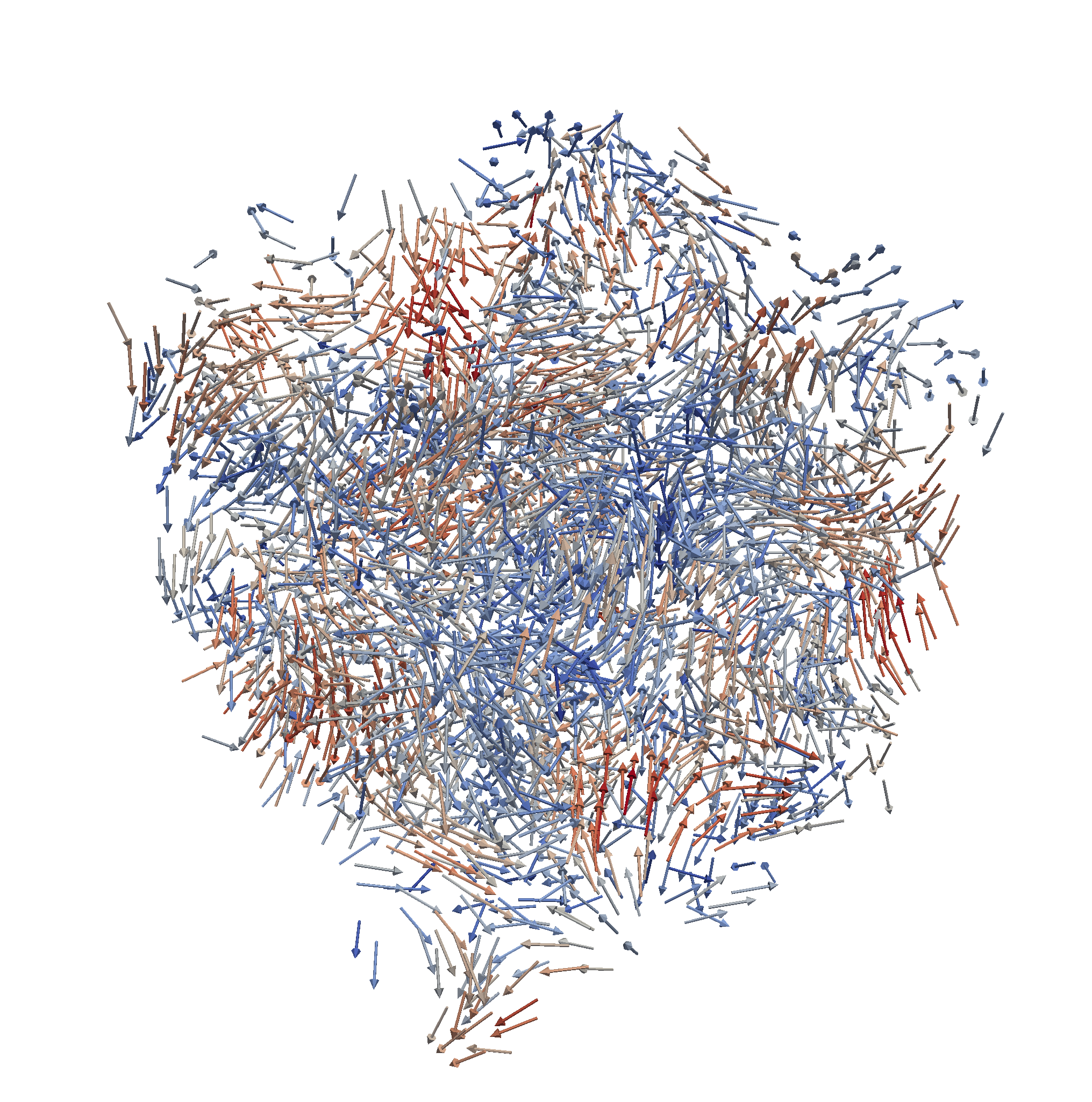}
\par\end{centering}
\label{fig:flow3D}\caption{The three dimensional velocity field corresponding to a single icosahedron-dodacoheron
compound in Fourier space, with random phases for the Fourier coefficients. }

\end{figure}

The three dimensional velocity field in real space implied by a $k$-space
discretization using a set of vertices (assuming the list of vertices
contain their reflections):

\begin{equation}
\mathbf{u}\left(\mathbf{x}\right)=\sum_{n}u_{n}^{i}e^{i\mathbf{k}_{n}\cdot\mathbf{x}}\hat{\mathbf{x}}_{i}\label{eq:real}
\end{equation}
where $\mathbf{\hat{x}}_{i}$ is the unit vector in the $i$th direction
(typically $x$, $y$ and $z$ directions). Now if we consider a single
unit icosahedron (i.e. of radius $k_{n}=1$), we can write the velocity
field as:
\begin{align}
\mathbf{u}_{ico}\left(\mathbf{x}\right) & =\sum_{\ell=1}^{6}\left(u_{\ell}^{i}e^{i\hat{\mathbf{k}}_{n\ell}\cdot\mathbf{x}}+c.c.\right)\hat{\mathbf{x}}_{i}\nonumber \\
 & =\sum_{\ell=1}^{6}\left(u_{\ell}^{+}e^{i\hat{\mathbf{k}}_{\ell}\cdot\mathbf{x}}\hat{\mathbf{h}}_{\ell}^{+}+u_{\ell}^{-}e^{i\hat{\mathbf{k}}_{\ell}\cdot\mathbf{x}}\hat{\mathbf{h}}_{\ell}^{-}+c.c.\right)\label{eq:ico1}
\end{align}
which is defined by the 12 complex coefficients $u_{\ell}^{\pm}$.
These coefficients correspond to weights of right and left handed
helicities in 6 different directions that are defined by the icosahedron.
 The same can be done for a unit dodecahedron:
\[
\mathbf{u}_{dod}\left(\mathbf{x}\right)=\sum_{\ell=1}^{10}\left(u_{\ell}^{+}e^{i\hat{\mathbf{k}}_{\ell}\cdot\mathbf{x}}\hat{\mathbf{h}}_{\ell}^{+}+u_{\ell}^{-}e^{i\hat{\mathbf{k}}_{\ell}\cdot\mathbf{x}}\hat{\mathbf{h}}_{\ell}^{-}+c.c.\right)
\]
\label{eq:dod1}In order to see the real space structure of such a
flow, we have considered an icosahedron dodecahedron compound with
randomly picked values for the coefficients $u_{\ell}^{\pm}$, and
constructed the flow using (\ref{eq:ico1}) and (\ref{eq:dod1}),
and plotted the result in Fig \ref{fig:flow3D}.

\subsection{Transition to two dimensions}

There are various limiting cases, such as rotating turbulence, MHD
with a strong mean magnetic field, turbulence in a thin film etc.
where the turbulent dynamics become two dimensional. However there
are some peculiar aspects of two dimensional turbulence and other
peculiar aspects of 2D shell models (or logarithmically discretized
models). Two dimensional turbulence is generally believed to result
in a dual cascade, where the enstrophy cascades in the forward and
energy cascades in the backward directions. However, logarithmic discretization
of 2D space leads to a numerical inconvenience that the equipartition
of energy between shells dominate over the inverse energy cascade.
Thus, 2D shell models can not describe the inverse energy cascade
(unless there are additional terms in the equation that keeps the
system away from this tendency to equipartition). Interestingly this
is less of an issue in three dimensional models. When the model is
isotropic, however, it is impossible to study the transition from
3D to 2D, since all the directions are, by construction, the same.
However, a nested polyhedra model can in principle be scaled in one
direction (say the $k_{z}$ direction) and spherical shells can be
flattened to pancake-like forms and therefore the transition to 2D
can be studied.

The limiting form of 2D turbulence can be obtained by simply setting
$k_{z}=0$ in the nested-polyhedra structure of the grid. However
an interesting phenomenon takes place in this limit. The projection
of the nodes $1$ to $5$ of the $n$th icosahedron (which gives a
regular pentagon) is the same as those of the nodes $10$ to $14$
of the $n+1$th and $15$ to $19$ of the $n-1$th dodecahedra. Similarly
the nodes $7$ to $11$ of the $n$ th icosahedron is the the same
as the nodes $0$ to $4$ of the $n+1$th and $5$ to $9$ of the
$n-1$th dodecahedra. This means that the $n$th ``shell'' (really
a circular ring or annulus) in a 2D model corresponds to at least
three different ``shells'' $n$, $n+1$ and $n-1$ in a 3D model.
As they fall exactly on the same points on the 2D space and they are
indistiguishable and lead to degeneracy.

Note that $\alpha=\arcsin\left(\frac{2}{3\lambda g}\right)$ is equivalent
to the form given in Table \ref{tab:angles} for $\alpha$. The smaller
pentagon which results from the projection of the inner points of
a dodacohedron on the $k_{z}=0$ plane has a circumscribed circle
of radius: 
\[
k_{\perp}=\frac{2k}{3\lambda g}\quad\text{.}
\]
The larger pentagon which comes from the projection of the outer points
of the dodecahedron and that which comes from those of its dual icosahedron
has a circumscribed circle of radius:
\[
k_{\perp}=\frac{2gk}{3\lambda}\quad\text{.}
\]
This gives a scaling factor of $g^{2}$ between two consecutive circles.
In other words, the wavenumbers of the shells can now be defined with:
\[
k_{n\perp}=k_{0\perp}\varphi^{n}
\]
 where $\varphi=g^{2}=\left(1+\sqrt{5}\right)/2$ and $k_{0\perp}=\frac{2k_{0}}{3\lambda g}=\frac{2\sqrt{2}}{\sqrt{3\left(5+\sqrt{5}\right)}}k_{0}$. 

The equations on the shells can be obtained by projection also. The
helicity directions become: 
\[
\hat{\mathbf{h}}_{\ell}^{\pm}=\hat{\mathbf{z}}\pm i\hat{\mathbf{k}}_{\perp}\times\hat{\mathbf{z}}
\]
which allows us to write seperate equations for $u_{ico}^{\pm}$ and
$u_{dod}^{\pm}$ for each scale simply by projecting the corresponding
three dimensional equations to two dimensions. This means that now,
we need to solve $4\times10$ equations for each scale, or noting
that half of these points are simply reflections of the rest of the
points, $4\times5=20$ equations for each shell.

The 4-fold degeneracy which appear in going from 3D to 2D is remarkable,
since in the standard 2D formulation one would consider only vorticity
(not helicities of both signs) and only one kind of polygon as a representation
of the 2D $k$-space\cite{gurcan:16a}. The effort discussed in this
work for going to a 2D formulation may be worthwhile however due to
the issue of unphysical shell equipartition overwhelming the inverse
cascade in 2D logarithmic discretization. A model which has the same
topological structure as the 3D one, may actually be able to reproduce
the $k^{-5/3}$ inverse cascade spectrum without requiring the expensive
hierarchical tree approach\cite{aurell:94,aurell:97}. However, this
particular issue is not the focus of the current article, and therefore
the concentrated effort neccessary for establishing the implications
of such a model is left to a future publication.

\section{Numerical Results}

The model, solves for all three components of the velocity field,
with an interaction matrix $M_{n\ell}^{\kappa ij}$ representing the
Navier-Stokes equation. In order to implement it, an object oriented
approach can be used, where each node has a list of its connecting
pairs, as can be inferred from Tables \ref{tab:idd} and \ref{tab:dii}
(as seen in ure \ref{fig:network}), and so that a sum over these
pairs can be computed rapidly. The resulting model is a stiff set
of ordinary differential equations (ODEs) on an exponantially coarse
grid, somewhat similar to the 2D model discussed in Ref. \cite{gurcan:16a}.
\begin{figure}
\begin{centering}
\includegraphics[width=0.49\columnwidth]{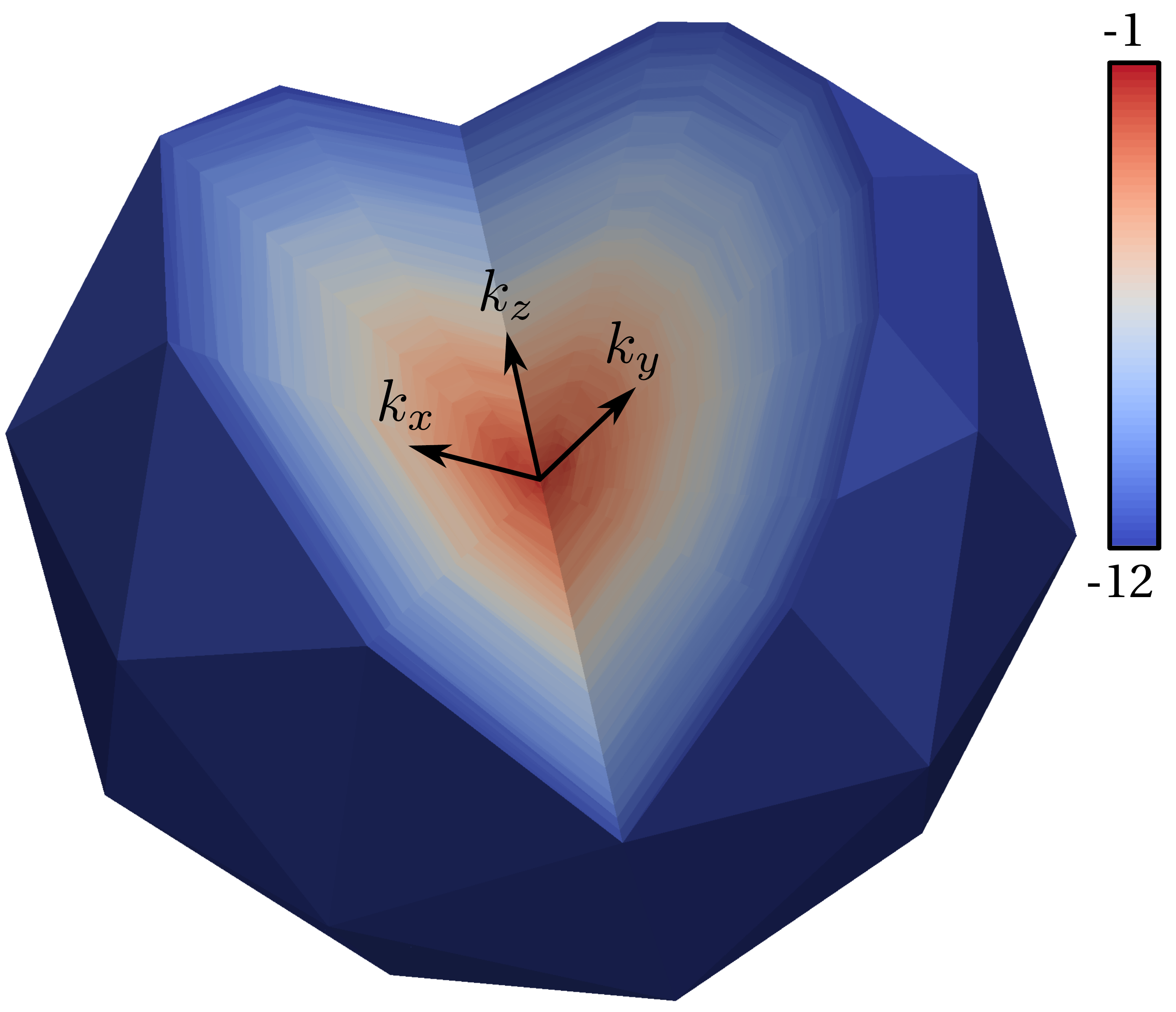}\includegraphics[width=0.49\columnwidth]{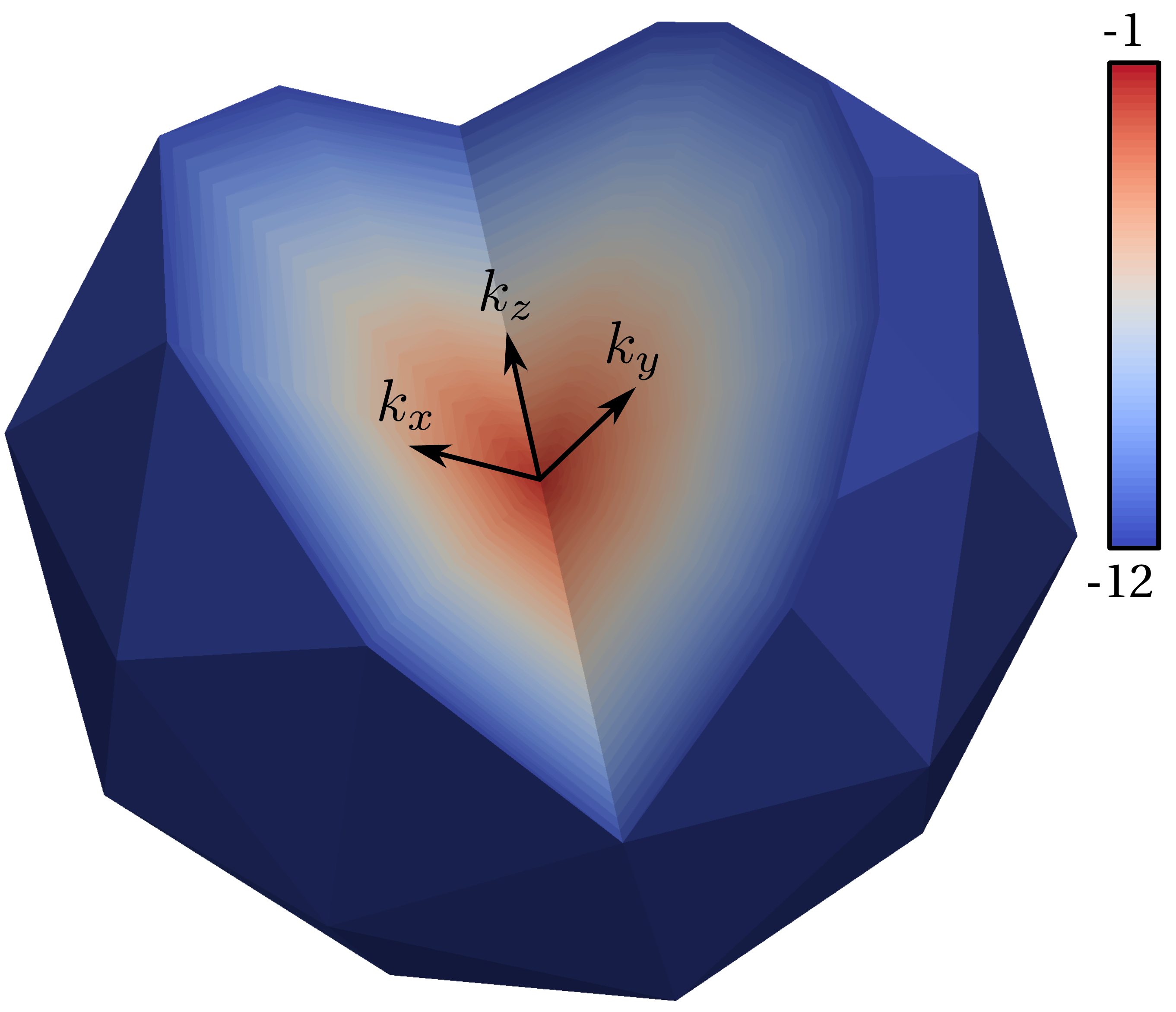}
\par\end{centering}
\caption{\label{fig:res3D}The resulting instantaneous 3D $k$-spectrum at
$t=250$ on the left and averaged over the range $t=[200,250]$ on
the right for the run with $N=60$ and $\nu=10^{-10}$. Here we used
a spherical log-log representation, where $E\left(k_{n}\right)=\log\left[\left(u_{n\ell x}^{2}+u_{n\ell y}^{2}+u_{n\ell z}^{2}\right)/k_{n}\right]$
is plotted with respect to $\boldsymbol{\kappa}_{n\ell}=\log\left(k_{n}\right)\hat{\mathbf{k}}_{\ell}$.
The resulting spectrum is consistent with the Kolmogorov spectrum
$E\left(k\right)\propto k^{-5/3}$ as shown in figure \ref{fig:res_spec}.}
\end{figure}
\begin{figure}
\begin{centering}
\includegraphics[width=0.98\columnwidth]{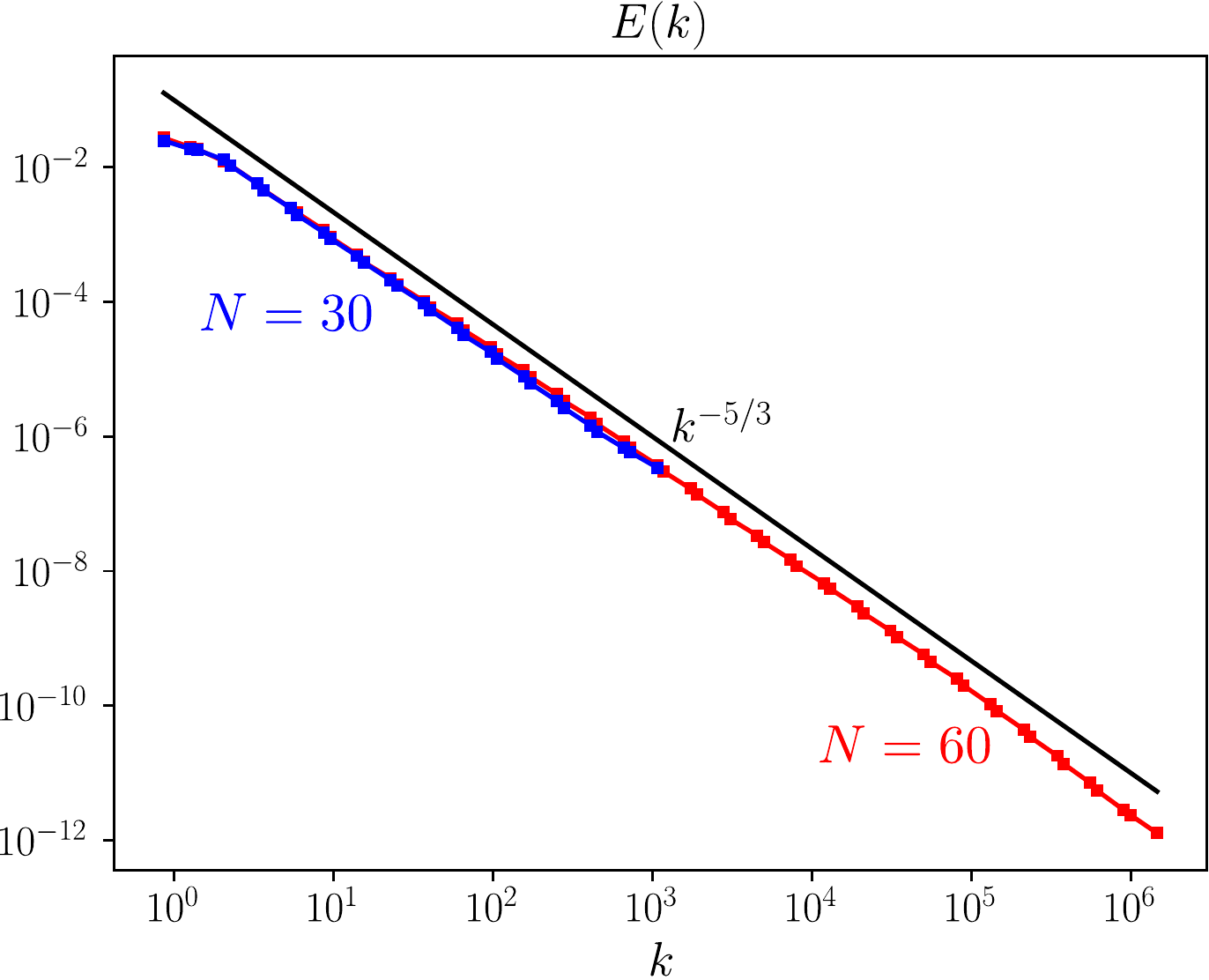}
\par\end{centering}
\caption{\label{fig:res_spec}Log-log plot of the spectral energy density $E\left(k\right)=\frac{1}{N_{\ell}k_{n}}\sum_{\ell,i}\left|u_{n\ell}^{i}\right|^{2}$
as a function of $k=k_{n}$. The blue (if in color) curve that spans
from $1$ to $10^{3}$ is a run with $N=30$ and $\nu=10^{-6}$ averaged
in the range $t=[800,1000]$, whereas the red curve (if in color)
that spans all the way up to $10^{6}$ is a run at the limit of currently
available resolution with $N=60$ and $\nu=10^{-10}$ averaged in
the range $t=[200,250]$.}
\end{figure}
\begin{figure}
\begin{centering}
\includegraphics[width=0.98\columnwidth]{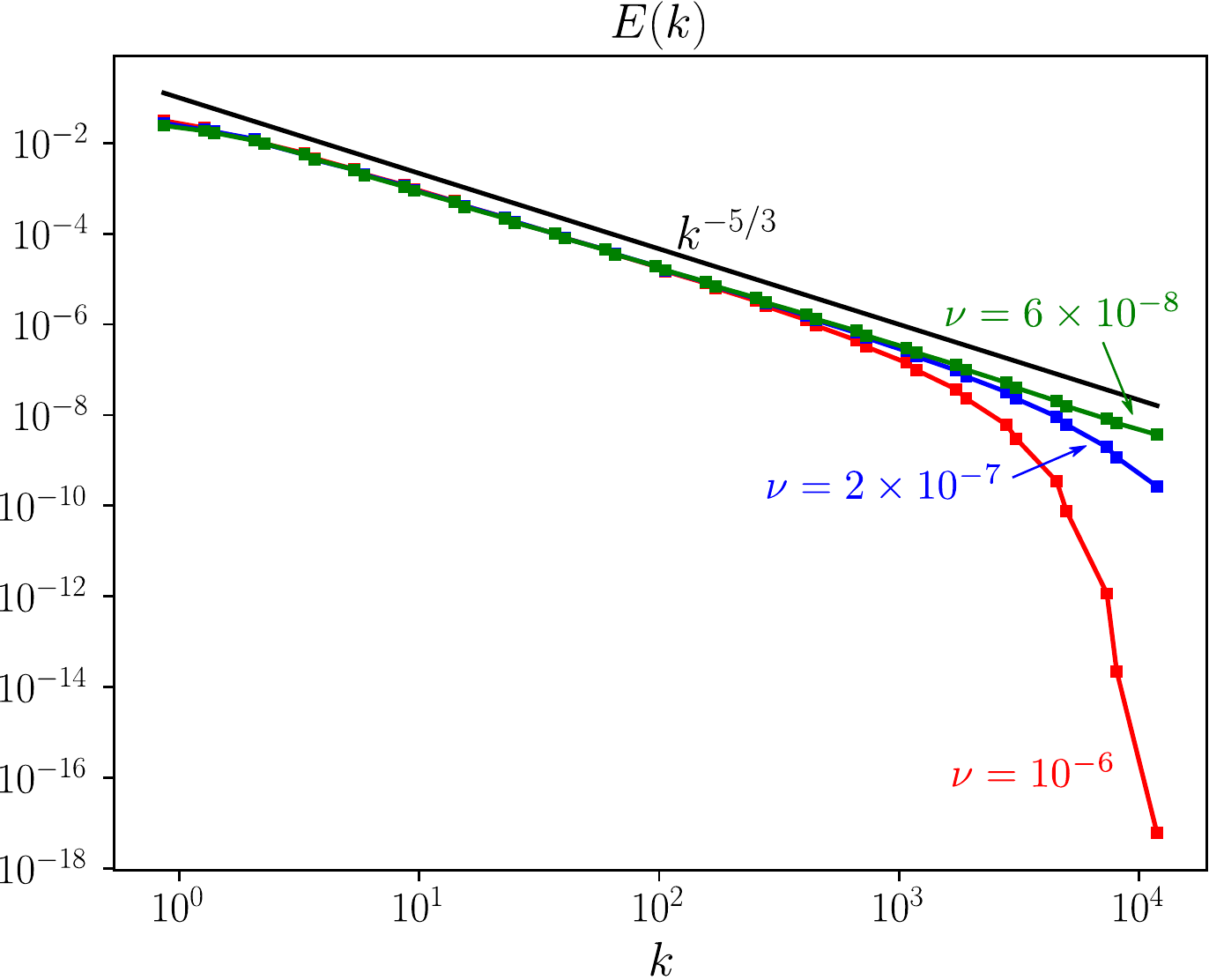}
\par\end{centering}
\caption{\label{fig:res_spec2}Log-log plot of spectral energy density as a
function of $k$ (see figure \ref{fig:res_spec} for definition) for
$N=40$. The blue (if in color) curve is with $\nu=2\times10^{-7}$,
while the red curve corresponds to $\nu=10^{-6}$, both averaged in
the range $t=[800,1000]$.}
\end{figure}

We have implemented the nested polyhedra model in Python with no parallelization,
which is distributed as an open source solver at \url{http://github.com/gurcani/nestp3d}.
It solves $3\times8\times N$ (where $N$ is the number of $k$-space
shells) complex system of equations, for the three components of the
velocity field for each half-polyhedra as discussed in section \ref{subsec:Further-simplifications}
above. We performed several runs ranging from $N=30$ to $N=60$. 

The three dimensional spectra for the highest resolution case (i.e.
$N=60$) are shown in figure \ref{fig:res3D}. As expected, with no
source of anisotropy, the resulting spectrum remains perfectly isotropic.
Nonetheless, it shows the capability of the model to resolve three
dimensional anisotropy accross many decades. The results of the two
limiting cases $N=30$ with $\nu=10^{-6}$ and $N=60$ with $\nu=10^{-10}$
are shown in figure \ref{fig:res_spec}, where the case $N=60$ (which
takes several weeks to compute on a pc workstation) covers almost
6 decades in $k$-space and has almost no need of a dissipative range
to reach steady state. This seems to be a feature of the model, which
may facilitate development of large eddy simulation (LES) versions
of itself. As can be seen in figure \ref{fig:res_spec}, the truncation
from 60 to 30 shells while varying $\nu$ accordingly, has almost
no effect on part of the spectrum that is resolved by the $N=30$
run. In order to study the effect of the existence of a dissipation
range, we have also varied the viscosity coefficient $\nu$. The results
for $\nu=10^{-6}$, $\nu=2\times10^{-7}$ and $\nu=6\times10^{-8}$
are considered for $N=40$, and shown in figure \ref{fig:res_spec2}.
As expected, increasing $\nu$ causes a dissipative range to appear,
but it doesn't change neither the saturation level, nor the slope
of the spectrum.

\begin{figure}
\begin{centering}
\includegraphics[width=0.98\columnwidth]{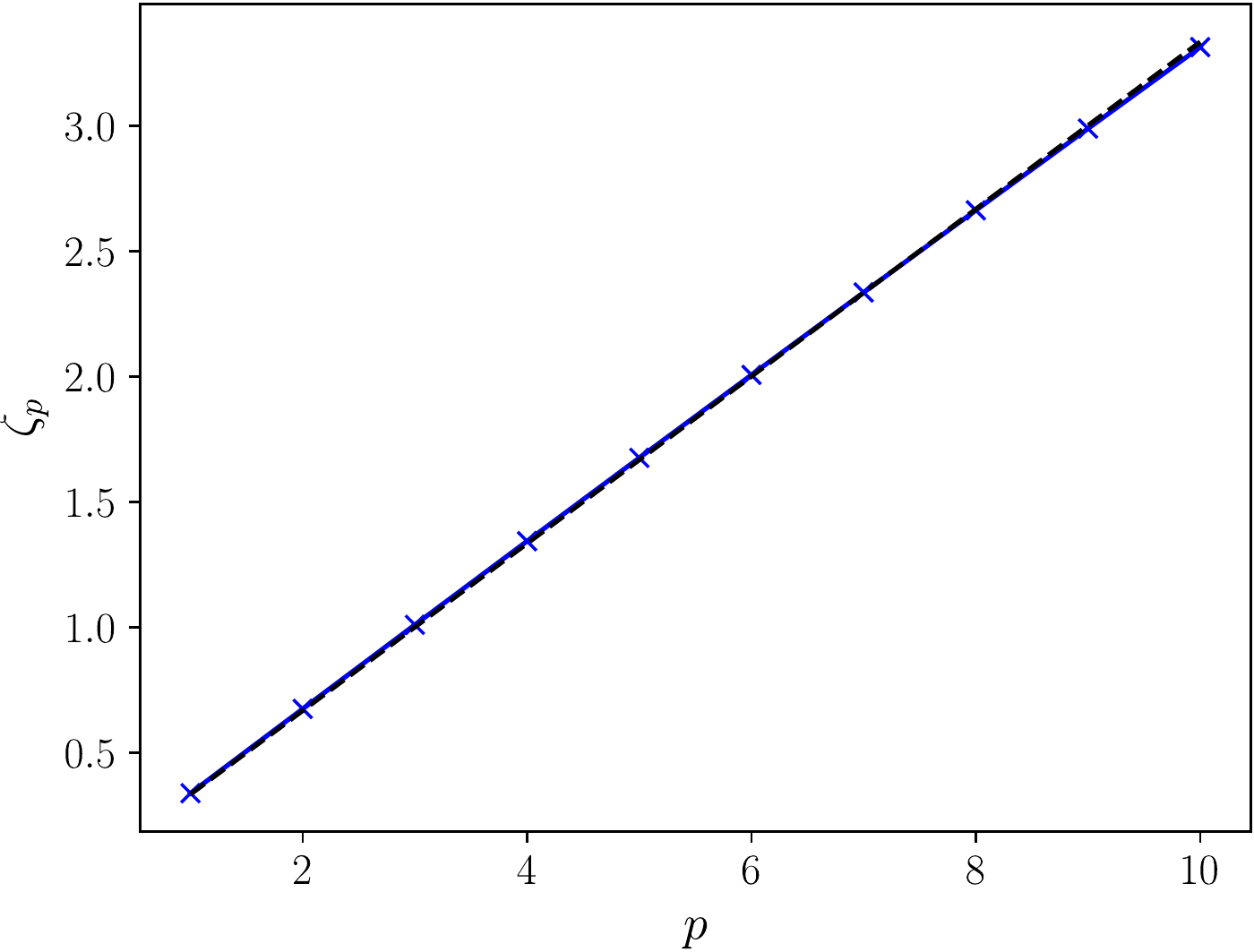}
\par\end{centering}
\caption{\label{fig:inter}The index $\zeta_{p}$ of the power law for the
structure function of order $p$ {[}i.e. $S_{p}\left(k_{n}\right)\sim k^{-\zeta_{p}}${]}
as a function of $p$, displaying a clear $S_{p}\left(k_{n}\right)\sim k_{n}^{-p/3}$
scaling. The case that is shown here corresponds to $N=60$, with
$\nu\sim10^{-10}$, which was averaged only from $t=200$ to $t=250$
over shells $N=4$ to $N=50$. However another case with $N=30$ was
integrated up to $t=25000$ so that an average could be computed from
$t=1000$ to $t=25000$ over shells $4$ to $12$, and it gives virtually
the same result.}
\end{figure}

Finally, we have performed some preliminary studies of intermittency
using this model, whose results are shown in figure \ref{fig:inter}.
Curiously, the model shows no sign of intermittency as it follows
the $S_{p}\left(k_{n}\right)\sim k_{n}^{-p/3}$ scaling in the inertial
range (i.e. $S_{p}\left(k_{n}\right)=\left\langle \frac{1}{N_{\ell}}\sum_{\ell}\left(\sum_{i}\left|u_{n\ell}^{i}\right|^{2}\right)^{p/2}\right\rangle $
where $\left\langle \cdot\right\rangle $denotes average over time).
On one hand, this is surprising, since for instance the GOY model,
which can be derived from the nested polyhedra model by additional
assumptions about the way the phases are organised, displays dynamical
multiscaling and intermittency\cite{pisarenko:93,benzi:96}, while
on the other hand, it is natural, since the model is perfectly self-similar
and made of a single three dimensional ``fractal''. In order to
further verify that our method of obtaining intermittency is valid,
we double-checked our results by repeating the exercise of Ref. \cite{pisarenko:93}
with standard GOY model as well as a model with alternating shells
(i.e. $k_{n}=k_{0}g^{n}$ for even $n$ and $k_{n}=k_{0}g^{n}\lambda$
for odd n), and found that while their results are rather robust for
shell models, our result of ``no intermittency'' is similarly robust
for our model. We repeated this exercise many times, and remain puzzled
by these results, since it seems to us paradoxal that the reduction
of a reduction can recover a property of the original system that
was lost in the first level of reduction. One possible explanation
is the fact that shell models do not rely on a single type of triad,
but is a result of a net transfer of conserved quantity from shell
to shell, where many similar triads may play a role. This sense of
``net flux'' computed over a set of similar triads may be the reason
why these models can somehow have intermittent dynamics, while our
model, which relies on a single triad family accross many scales,
does not.

\section{Conclusion}

A dodecahedron-icosahedron compound discretization of the Navier-Stokes
equation, which is proposed in this paper, gives the expected wave-number
spectrum $k^{-5/3}$ of Kolmogorov and can possibly be used to study
the spectra in three dimensional fluid turbulence. The advantage of
a formulation based on logarithmic scaling with a constant number
of nodes per scale is that a very large inertial range (i.e. very
high Reynolds numbers) can be considered with a much lower number
of degrees of freedom. Further simplifications of the model were proposed
using the symmetry in $k$-space and the helical decomposition. It
was shown that, when isotropy and a particular relation between phases
are imposed, the nested polyhedra model reduces to a GOY model. This
provides an actual ``systematic derivation'' of the latter, since
the nested polyhedra model itself was obtained by a systematic reduction/decimation
of the continuous wave-number space to a finite set of self-similar
triads. The straightforward issue of reconstruction of the velocity
field from the nested polyhedra description is also discussed in order
to demonstrate the richness of the types of flows that the highly
reduced model can sustain. Transition to two dimensions is briefly
discussed as a reference for future work. It is noteworthy that one
can derive a model for describing two dimensional turbulence, which
has the same topological network structure of the three dimensional
one. Preliminary studies show that the model as described in this
paper, show no sign of intermittence since it follows the $S_{p}\left(k_{n}\right)\sim k_{n}^{-p/3}$
scaling in the inertial range. This is curious, since the GOY model,
which can be derived from the nested polyhedra model by additional
assumptions about the way the phases are organised, displays dynamical
multiscaling and intermittency\cite{pisarenko:93,benzi:96}. On the
other hand, it is natural, since the model is perfectly self-similar
and made of a single three dimensional ``fractal''.

Note that the icosahedron and the dodecahedron (i.e. its dual), together
forms a compound polyhedron called ``dodecahedron-icosahedron compound''
(i.e. Wenninger model index 47 \cite{wenninger:book:74}). The nested
polyhedra model that we introduced here can be seen as a discretization
of the $k$-space using these objects. A faceting of this compound
polyhedron is a Catalan solid called ``rhombic triacontahedron'',
which is also the dual of an Archimedean solid called ``icosidodecahedron''.
One could use this connection to ``refine'' the k-space, that is
divided in self-similar nested polyhedra, by introducing an icosidodecahedron
between two dodecahedron-icosahedron compounds that constitute the
nested polyhedra model. The resulting model would use nested ``icosidodecahedron-rhombic
triacontahedron compounds'' as the building blocks for the nested
polyhedra model. It would be interesing to determine, using wavenumber
triad matching conditions, if there exist more of these complex polyhedra,
which can be used to develop more and more complex nested polyhedra
models. One could then speculate that if one would initialize the
turbulence on an icosahedron in $k$-space, the full system of Navier
stokes equations (with no truncation) would result in the turbulence
energy going from one type of complex polyhedron to another, without
ever leaving the space of compound polyhedra. This is remarkable as
it would transform the infinite system to a set of nested polyhedra.
Incidentally, the use of multiple types of polyhedra, would also introduce
multifractality naturally.

When the approach detailed in this study is applied to a system that
supports waves (i.e. linearly), the resulting network model can be
suitable for the study of various different phenomena including synchronization\cite{kuramoto:book:1984},
small world or scale freedom\cite{wang:03} since it can be thought
of as a complex network of coupled ``oscillators'' \cite{strogatz:01}.
The structure of the model is somewhat similar to those describing
food web networks \cite{pimm:91} or supply chains, which implies
a complex adaptive system. Such analogies are used regularly in plasma
turbulence, especially in the presence of large scale flow structures
called zonal flows\cite{diamond:05,gurcan:15}. It would be interesting
to see if a similar analogy can be extended to fluid turbulence and
to what extent a rigorous mathematical relation can be established
between such reduced models of turbulence and analogous ones from
biology and other fields.
\begin{acknowledgments}
The author would like to thank P. Morel, P. H. Diamond, R. Grappin
and attendants and the organizers of the \emph{Festival de Théorie,
Aix en Provence} in 2015.
\end{acknowledgments}


\begin{thebibliography}{32}
\expandafter\ifx\csname natexlab\endcsname\relax\def\natexlab#1{#1}\fi
\expandafter\ifx\csname bibnamefont\endcsname\relax
  \def\bibnamefont#1{#1}\fi
\expandafter\ifx\csname bibfnamefont\endcsname\relax
  \def\bibfnamefont#1{#1}\fi
\expandafter\ifx\csname citenamefont\endcsname\relax
  \def\citenamefont#1{#1}\fi
\expandafter\ifx\csname url\endcsname\relax
  \def\url#1{\texttt{#1}}\fi
\expandafter\ifx\csname urlprefix\endcsname\relax\def\urlprefix{URL }\fi
\providecommand{\bibinfo}[2]{#2}
\providecommand{\eprint}[2][]{\url{#2}}

\bibitem[{\citenamefont{Falkovich}(2009)}]{falkovich:09}
\bibinfo{author}{\bibfnamefont{G.}~\bibnamefont{Falkovich}},
  \bibinfo{journal}{J. Phys. A} \textbf{\bibinfo{volume}{42}},
  \bibinfo{pages}{123001} (\bibinfo{year}{2009}).

\bibitem[{\citenamefont{Frisch}(1995)}]{frisch}
\bibinfo{author}{\bibfnamefont{U.}~\bibnamefont{Frisch}},
  \emph{\bibinfo{title}{Turbulence: The Legacy of A. N. Kolmogorov}}
  (\bibinfo{publisher}{Cambridge University Press},
  \bibinfo{address}{Cambridge}, \bibinfo{year}{1995}).

\bibitem[{\citenamefont{Galtier et~al.}(2000)\citenamefont{Galtier, Nazarenko,
  Newell, and Pouquet}}]{galtier:00}
\bibinfo{author}{\bibfnamefont{S.}~\bibnamefont{Galtier}},
  \bibinfo{author}{\bibfnamefont{S.~V.} \bibnamefont{Nazarenko}},
  \bibinfo{author}{\bibfnamefont{A.~C.} \bibnamefont{Newell}},
  \bibnamefont{and} \bibinfo{author}{\bibfnamefont{A.}~\bibnamefont{Pouquet}},
  \bibinfo{journal}{Journal of Plasma Physics} \textbf{\bibinfo{volume}{63}},
  \bibinfo{pages}{447} (\bibinfo{year}{2000}).

\bibitem[{\citenamefont{Leith}(1971)}]{leith:71}
\bibinfo{author}{\bibfnamefont{C.~E.} \bibnamefont{Leith}},
  \bibinfo{journal}{J. Atmos. Sci.} \textbf{\bibinfo{volume}{28}},
  \bibinfo{pages}{145} (\bibinfo{year}{1971}).

\bibitem[{\citenamefont{Bowman}(1996)}]{bowman:96}
\bibinfo{author}{\bibfnamefont{J.~C.} \bibnamefont{Bowman}},
  \bibinfo{journal}{Journal of Scientific Computing}
  \textbf{\bibinfo{volume}{11}}, \bibinfo{pages}{343} (\bibinfo{year}{1996}).

\bibitem[{\citenamefont{Meldi and Sagaut}(2012)}]{meldi:12}
\bibinfo{author}{\bibfnamefont{M.}~\bibnamefont{Meldi}} \bibnamefont{and}
  \bibinfo{author}{\bibfnamefont{P.}~\bibnamefont{Sagaut}},
  \bibinfo{journal}{Journal of Fluid Mechanics} \textbf{\bibinfo{volume}{711}},
  \bibinfo{pages}{364} (\bibinfo{year}{2012}), ISSN \bibinfo{issn}{1469-7645}.

\bibitem[{\citenamefont{Lilly}(1989)}]{lilly:89}
\bibinfo{author}{\bibfnamefont{D.~K.} \bibnamefont{Lilly}},
  \bibinfo{journal}{J. Atmos. Sci.} \textbf{\bibinfo{volume}{46}},
  \bibinfo{pages}{2026} (\bibinfo{year}{1989}).

\bibitem[{\citenamefont{{L'vov} and Nazarenko}(2006)}]{lvov:06}
\bibinfo{author}{\bibfnamefont{V.~S.} \bibnamefont{{L'vov}}} \bibnamefont{and}
  \bibinfo{author}{\bibfnamefont{S.}~\bibnamefont{Nazarenko}},
  \bibinfo{journal}{{JETP Letters}} \textbf{\bibinfo{volume}{83}},
  \bibinfo{pages}{541} (\bibinfo{year}{2006}).

\bibitem[{\citenamefont{Thiagalingam and Sagaut}(2012)}]{thiagalinam:12}
\bibinfo{author}{\bibfnamefont{I.}~\bibnamefont{Thiagalingam}}
  \bibnamefont{and} \bibinfo{author}{\bibfnamefont{P.}~\bibnamefont{Sagaut}},
  \bibinfo{journal}{Physics of Fluids} \textbf{\bibinfo{volume}{24}},
  \bibinfo{eid}{115109} (\bibinfo{year}{2012}).

\bibitem[{\citenamefont{Biferale}(2003)}]{biferale:03}
\bibinfo{author}{\bibfnamefont{L.}~\bibnamefont{Biferale}},
  \bibinfo{journal}{Ann. Rev. Fluid Mech.} \textbf{\bibinfo{volume}{35}},
  \bibinfo{pages}{441} (\bibinfo{year}{2003}).

\bibitem[{\citenamefont{G\"urcan et~al.}(2016)\citenamefont{G\"urcan, Morel,
  Kobayashi, Singh, Xu, and Diamond}}]{gurcan:16a}
\bibinfo{author}{\bibfnamefont{{\"O}.~D.} \bibnamefont{G\"urcan}},
  \bibinfo{author}{\bibfnamefont{P.}~\bibnamefont{Morel}},
  \bibinfo{author}{\bibfnamefont{S.}~\bibnamefont{Kobayashi}},
  \bibinfo{author}{\bibfnamefont{R.}~\bibnamefont{Singh}},
  \bibinfo{author}{\bibfnamefont{S.}~\bibnamefont{Xu}}, \bibnamefont{and}
  \bibinfo{author}{\bibfnamefont{P.~H.} \bibnamefont{Diamond}},
  \bibinfo{journal}{Phys. Rev. E} \textbf{\bibinfo{volume}{94}},
  \bibinfo{pages}{033106} (\bibinfo{year}{2016}).

\bibitem[{\citenamefont{Zeitlin}(1991)}]{zeitlin:91}
\bibinfo{author}{\bibfnamefont{V.}~\bibnamefont{Zeitlin}},
  \bibinfo{journal}{Physica D: Nonlinear Phenomena}
  \textbf{\bibinfo{volume}{49}}, \bibinfo{pages}{353 } (\bibinfo{year}{1991}),
  ISSN \bibinfo{issn}{0167-2789}.

\bibitem[{\citenamefont{Kerr and Siggia}(1978)}]{kerr:78}
\bibinfo{author}{\bibfnamefont{R.~M.} \bibnamefont{Kerr}} \bibnamefont{and}
  \bibinfo{author}{\bibfnamefont{E.~D.} \bibnamefont{Siggia}},
  \bibinfo{journal}{Journal of Statistical Physics}
  \textbf{\bibinfo{volume}{19}}, \bibinfo{pages}{543} (\bibinfo{year}{1978}),
  ISSN \bibinfo{issn}{1572-9613},
  \urlprefix\url{http://dx.doi.org/10.1007/BF01011698}.

\bibitem[{\citenamefont{Eggers and Grossmann}(1991)}]{eggers:91}
\bibinfo{author}{\bibfnamefont{J.}~\bibnamefont{Eggers}} \bibnamefont{and}
  \bibinfo{author}{\bibfnamefont{S.}~\bibnamefont{Grossmann}},
  \bibinfo{journal}{Physics of Fluids A: Fluid Dynamics}
  \textbf{\bibinfo{volume}{3}}, \bibinfo{pages}{1958} (\bibinfo{year}{1991}),
  \urlprefix\url{http://dx.doi.org/10.1063/1.857926}.

\bibitem[{\citenamefont{V\'azquez-Semadeni and
  Scalo}(1992)}]{vazquez-semadeni:92}
\bibinfo{author}{\bibfnamefont{E.}~\bibnamefont{V\'azquez-Semadeni}}
  \bibnamefont{and} \bibinfo{author}{\bibfnamefont{J.}~\bibnamefont{Scalo}},
  \bibinfo{journal}{Phys. Rev. Lett.} \textbf{\bibinfo{volume}{68}},
  \bibinfo{pages}{2921} (\bibinfo{year}{1992}).

\bibitem[{\citenamefont{Frisch et~al.}(1978)\citenamefont{Frisch, Sulem, and
  Nelkin}}]{frisch:78}
\bibinfo{author}{\bibfnamefont{U.}~\bibnamefont{Frisch}},
  \bibinfo{author}{\bibfnamefont{P.-L.} \bibnamefont{Sulem}}, \bibnamefont{and}
  \bibinfo{author}{\bibfnamefont{M.}~\bibnamefont{Nelkin}},
  \bibinfo{journal}{Journal of Fluid Mechanics} \textbf{\bibinfo{volume}{87}},
  \bibinfo{pages}{719} (\bibinfo{year}{1978}), ISSN \bibinfo{issn}{1469-7645}.

\bibitem[{\citenamefont{Biferale et~al.}(1995)\citenamefont{Biferale, Lambert,
  Lima, and Paladin}}]{biferale:95}
\bibinfo{author}{\bibfnamefont{L.}~\bibnamefont{Biferale}},
  \bibinfo{author}{\bibfnamefont{A.}~\bibnamefont{Lambert}},
  \bibinfo{author}{\bibfnamefont{R.}~\bibnamefont{Lima}}, \bibnamefont{and}
  \bibinfo{author}{\bibfnamefont{G.}~\bibnamefont{Paladin}},
  \bibinfo{journal}{Physica D: Nonlinear Phenomena}
  \textbf{\bibinfo{volume}{80}}, \bibinfo{pages}{105 } (\bibinfo{year}{1995}),
  ISSN \bibinfo{issn}{0167-2789}.

\bibitem[{\citenamefont{L\char39{}vov et~al.}(1998)\citenamefont{L\char39{}vov,
  Podivilov, Pomyalov, Procaccia, and Vandembroucq}}]{lvov:98}
\bibinfo{author}{\bibfnamefont{V.~S.} \bibnamefont{L\char39{}vov}},
  \bibinfo{author}{\bibfnamefont{E.}~\bibnamefont{Podivilov}},
  \bibinfo{author}{\bibfnamefont{A.}~\bibnamefont{Pomyalov}},
  \bibinfo{author}{\bibfnamefont{I.}~\bibnamefont{Procaccia}},
  \bibnamefont{and}
  \bibinfo{author}{\bibfnamefont{D.}~\bibnamefont{Vandembroucq}},
  \bibinfo{journal}{Phys. Rev. E} \textbf{\bibinfo{volume}{58}},
  \bibinfo{pages}{1811} (\bibinfo{year}{1998}).

\bibitem[{\citenamefont{Ohkitani and Yamada}(1989)}]{ohkitani:89}
\bibinfo{author}{\bibfnamefont{K.}~\bibnamefont{Ohkitani}} \bibnamefont{and}
  \bibinfo{author}{\bibfnamefont{M.}~\bibnamefont{Yamada}},
  \bibinfo{journal}{Progress of Theoretical Physics}
  \textbf{\bibinfo{volume}{81}}, \bibinfo{pages}{329} (\bibinfo{year}{1989}).

\bibitem[{\citenamefont{Jensen et~al.}(1991)\citenamefont{Jensen, Paladin, and
  Vulpiani}}]{jensen:91}
\bibinfo{author}{\bibfnamefont{M.~H.} \bibnamefont{Jensen}},
  \bibinfo{author}{\bibfnamefont{G.}~\bibnamefont{Paladin}}, \bibnamefont{and}
  \bibinfo{author}{\bibfnamefont{A.}~\bibnamefont{Vulpiani}},
  \bibinfo{journal}{Phys. Rev. A} \textbf{\bibinfo{volume}{43}},
  \bibinfo{pages}{798} (\bibinfo{year}{1991}).

\bibitem[{\citenamefont{Pisarenko et~al.}(1993)\citenamefont{Pisarenko,
  Biferale, Courvoisier, Frisch, and Vergassola}}]{pisarenko:93}
\bibinfo{author}{\bibfnamefont{D.}~\bibnamefont{Pisarenko}},
  \bibinfo{author}{\bibfnamefont{L.}~\bibnamefont{Biferale}},
  \bibinfo{author}{\bibfnamefont{D.}~\bibnamefont{Courvoisier}},
  \bibinfo{author}{\bibfnamefont{U.}~\bibnamefont{Frisch}}, \bibnamefont{and}
  \bibinfo{author}{\bibfnamefont{M.}~\bibnamefont{Vergassola}},
  \bibinfo{journal}{Physics of Fluids A} \textbf{\bibinfo{volume}{5}},
  \bibinfo{pages}{2533} (\bibinfo{year}{1993}).

\bibitem[{\citenamefont{Wenninger}(1974)}]{wenninger:book:74}
\bibinfo{author}{\bibfnamefont{M.~J.} \bibnamefont{Wenninger}},
  \emph{\bibinfo{title}{Polyhedron Models}} (\bibinfo{publisher}{Cambridge
  University Press}, \bibinfo{address}{Cambridge}, \bibinfo{year}{1974}).

\bibitem[{\citenamefont{De~Pietro et~al.}(2015)\citenamefont{De~Pietro,
  Biferale, and Mailybaev}}]{depietro:15}
\bibinfo{author}{\bibfnamefont{M.}~\bibnamefont{De~Pietro}},
  \bibinfo{author}{\bibfnamefont{L.}~\bibnamefont{Biferale}}, \bibnamefont{and}
  \bibinfo{author}{\bibfnamefont{A.~A.} \bibnamefont{Mailybaev}},
  \bibinfo{journal}{Phys. Rev. E} \textbf{\bibinfo{volume}{92}},
  \bibinfo{pages}{043021} (\bibinfo{year}{2015}).

\bibitem[{\citenamefont{Aurell et~al.}(1994)\citenamefont{Aurell, Frick, and
  Shaidurov}}]{aurell:94}
\bibinfo{author}{\bibfnamefont{E.}~\bibnamefont{Aurell}},
  \bibinfo{author}{\bibfnamefont{P.}~\bibnamefont{Frick}}, \bibnamefont{and}
  \bibinfo{author}{\bibfnamefont{V.}~\bibnamefont{Shaidurov}},
  \bibinfo{journal}{Physica D: Nonlinear Phenomena}
  \textbf{\bibinfo{volume}{72}}, \bibinfo{pages}{95 } (\bibinfo{year}{1994}),
  ISSN \bibinfo{issn}{0167-2789}.

\bibitem[{\citenamefont{Aurell et~al.}(1997)\citenamefont{Aurell, Dormy, and
  Frick}}]{aurell:97}
\bibinfo{author}{\bibfnamefont{E.}~\bibnamefont{Aurell}},
  \bibinfo{author}{\bibfnamefont{E.}~\bibnamefont{Dormy}}, \bibnamefont{and}
  \bibinfo{author}{\bibfnamefont{P.}~\bibnamefont{Frick}},
  \bibinfo{journal}{Phys. Rev. E} \textbf{\bibinfo{volume}{56}},
  \bibinfo{pages}{1692} (\bibinfo{year}{1997}).

\bibitem[{\citenamefont{Benzi et~al.}(1996)\citenamefont{Benzi, Biferale, Kerr,
  and Trovatore}}]{benzi:96}
\bibinfo{author}{\bibfnamefont{R.}~\bibnamefont{Benzi}},
  \bibinfo{author}{\bibfnamefont{L.}~\bibnamefont{Biferale}},
  \bibinfo{author}{\bibfnamefont{R.~M.} \bibnamefont{Kerr}}, \bibnamefont{and}
  \bibinfo{author}{\bibfnamefont{E.}~\bibnamefont{Trovatore}},
  \bibinfo{journal}{Phys. Rev. E} \textbf{\bibinfo{volume}{53}},
  \bibinfo{pages}{3541} (\bibinfo{year}{1996}).

\bibitem[{\citenamefont{Kuramoto}(1984)}]{kuramoto:book:1984}
\bibinfo{author}{\bibfnamefont{Y.}~\bibnamefont{Kuramoto}},
  \emph{\bibinfo{title}{{Chemical Oscillations, Waves, and Turbulence}}}
  (\bibinfo{publisher}{Springer--Verlag}, \bibinfo{address}{New York},
  \bibinfo{year}{1984}).

\bibitem[{\citenamefont{Wang and Chen}(2003)}]{wang:03}
\bibinfo{author}{\bibfnamefont{X.~F.} \bibnamefont{Wang}} \bibnamefont{and}
  \bibinfo{author}{\bibfnamefont{G.}~\bibnamefont{Chen}},
  \bibinfo{journal}{IEEE Circuits and Systems Magazine}
  \textbf{\bibinfo{volume}{3}}, \bibinfo{pages}{6} (\bibinfo{year}{2003}).

\bibitem[{\citenamefont{Strogatz}(2001)}]{strogatz:01}
\bibinfo{author}{\bibfnamefont{S.~H.} \bibnamefont{Strogatz}},
  \bibinfo{journal}{Nature} \textbf{\bibinfo{volume}{410}},
  \bibinfo{pages}{268} (\bibinfo{year}{2001}).

\bibitem[{\citenamefont{Pimm et~al.}(1991)\citenamefont{Pimm, Lawton, and
  Cohen}}]{pimm:91}
\bibinfo{author}{\bibfnamefont{S.~L.} \bibnamefont{Pimm}},
  \bibinfo{author}{\bibfnamefont{J.~H.} \bibnamefont{Lawton}},
  \bibnamefont{and} \bibinfo{author}{\bibfnamefont{J.~E.} \bibnamefont{Cohen}},
  \bibinfo{journal}{Nature} \textbf{\bibinfo{volume}{350}},
  \bibinfo{pages}{669} (\bibinfo{year}{1991}).

\bibitem[{\citenamefont{Diamond et~al.}(2005)\citenamefont{Diamond,
  {S.-I.~Itoh}, Itoh, and Hahm}}]{diamond:05}
\bibinfo{author}{\bibfnamefont{P.~H.} \bibnamefont{Diamond}},
  \bibinfo{author}{\bibnamefont{{S.-I.~Itoh}}},
  \bibinfo{author}{\bibfnamefont{K.}~\bibnamefont{Itoh}}, \bibnamefont{and}
  \bibinfo{author}{\bibfnamefont{T.~S.} \bibnamefont{Hahm}},
  \bibinfo{journal}{PPCF} \textbf{\bibinfo{volume}{47}}, \bibinfo{pages}{R35}
  (\bibinfo{year}{2005}).

\bibitem[{\citenamefont{G\"{u}rcan and Diamond}(2015)}]{gurcan:15}
\bibinfo{author}{\bibfnamefont{{\"{O}}.~D.} \bibnamefont{G\"{u}rcan}}
  \bibnamefont{and} \bibinfo{author}{\bibfnamefont{P.}~\bibnamefont{Diamond}},
  \bibinfo{journal}{Journal of Physics A: Mathematical and Theoretical}
  \textbf{\bibinfo{volume}{48}}, \bibinfo{pages}{293001}
  (\bibinfo{year}{2015}).

\end{thebibliography}
\end{document}